\documentstyle[TSF11-ed,my]{article}
\begin{document}
\small
\title{\sffamily EXPLICIT-SCALES PROJECTIONS OF THE PARTITIONED
NON-LINEAR TERM IN DIRECT NUMERICAL SIMULATION OF THE NAVIER-STOKES
EQUATION}

\author{
\vv{\bfseries\sffamily
David McComb and Alistair Young}\\
\vv
Department of Physics and Astronomy\\
\vv
University of Edinburgh\\
\vv
James Clerk Maxwell Building\\
\vv
Mayfield Road\\
\vv
Edinburgh EH9 3JZ\\
\vv
United Kingdom}

\maketitle

\section{ABSTRACT}

In this paper we consider the properties of the internal partitions of
the nonlinear term, obtained when a filter with a sharp cutoff is introduced
in wavenumber space.  We see what appears to be some
degree of independence of the choice of the position of the
cutoff wavenumber for both instantaneous and time-integrated
partitioned nonlinearities.  We also investigate the basic idea of an
eddy-viscosity model for subgrid terms and have found that while
phase modelling will be very poor, amplitude modelling can be far more
successful.

\section{INTRODUCTION}

As is well known, full numerical simulation of any significant
turbulent flow lies far beyond the scope of current computational
resources,
the main problem being the large number of degrees of freedom
involved in the problem.  As these degrees of freedom may be represented
by, for instance, the number of independently excited modes in
wavenumber space, the problem becomes one of eliminating
modes, in some statistical sense, in order to bring the reduced
number of degrees of freedom within the capacity of current
(or even future) computers.  One such way by which we may
systematically obtain such a reduction in the number of modes is by
the use of a Renormalization Group (RG) calculation.  A general
account of the background to this work has been given in the review by
McComb (1995).

In this study, we are undertaking direct numerical simulations
(DNS) of homogeneous, isotropic, incompressible turbulence in
a box with periodic boundary conditions, in order to assess the underlying
feasibility of using RG to reduce the size of the computational
problem.  We have already reported some results on the
use of conditional averages (McComb {\sl et al.} 1997,
Machiels 1997) as previously formulated by
McComb {\sl et al} (1992) and McComb and Watt (1992).
In the present paper
we concentrate on the Hilbert space partitions of the
nonlinear terms and their filtered projections in order to
assess the appropriateness of the `eddy viscosity' concept.
Results of this study should have direct relevance to large
eddy simulations (LES) in general, as well as to RG.

\section{THE PARTITIONED NONLINEAR TERM}

Consider the forced Navier-Stokes equation for stationary turbulence,
\bea
\lefteqn{ \left( \ddt + \nu k^2 \right) u_\alpha(\Vec{k},t) } \nonumber \\
& = & M_{\alpha\beta\gamma}(\Vec{k}) \int d^3 j
u_\beta(\Vec{j},t) u_\gamma(\Vec{k}-\Vec{j},t) \nonumber \\
& + & f_\alpha(\Vec{k},t),
\label{NSEf}
\eea
where $\Vec{u}(\Vec{k},t)$ is the velocity field in Fourier-space,
$\nu$ is the kinematic viscosity, 
$M_{\alpha\beta\gamma}(\Vec{k})$ is given by
\be
\Mabg,
\ee
where
\be
\Dab
\ee
and $\Vec{f}(\Vec{k},t)$ is a forcing term used to achieve
stationarity.  We may rewrite
equation (\ref{NSEf}) in a highly symbolic form as
\be
L_0 u = M u u + f \label{L_0u=Muu}.
\ee

The nonlinear term ($Muu$ in our shorthand notation)
may be {\bf partitioned} by introducing a cutoff
at $k=k_1$ and defining $u^-$ and $u^+$ such that
$u_\alpha(\Vec{k},t) = u_\alpha^-(\Vec{k},t)$ for $0<k<k_1$
and $u_\alpha(\Vec{k},t) = u_\alpha^+(\Vec{k},t)$ for
$k_1<k<k_0$.  The {\sl maximum} cutoff wavenumber, $k_0$, is
of the same order of magnitude as the Kolmogorov dissipation
wavenumber and is
defined via the dissipation integral,
\be
\varepsilon
  =      \int_{0}^{\infty}\nu k^2 E(k) dk
  \simeq \int_{0}^{k_0}\nu k^2 E(k) dk
\end{equation}
where $\varepsilon$ is the dissipation rate.

Equation (\ref{L_0u=Muu}) can now be expanded to give
\be
L_0 u = \psi^{--} + \psi^{-+} + \psi^{++} + f \label{L0u=partition}
\ee
where the {\bf partitions} are defined by
\bea
\psi^{--} & = & M u^- u^- \\
\psi^{-+} & = & 2 M u^- u^+ \\
\psi^{++} & = & M u^+ u^+
\eea
and we further define
\be
\psi = \psi^{--} +\psi^{-+} + \psi^{++} = M u u.
\ee

We may now `solve' equation (\ref{L0u=partition}) by introducing
$G_0$ where, for some field $X(\Vec{k},t)$,
\bea
G_0 X(\Vec{k},t) & = & L_0^{-1} X(\Vec{k},t) \nonumber \\
& = & \int e^{-\nu k^2 ( t - t\p )}
X(\Vec{k},t\p) dt\p
\eea
so that
\be
u = \phi^{--} + \phi^{-+} + \phi^{++} + G_0 f, \label{u=partition}
\ee
where
\bea
\phi^{--} & = & G_0 \psi^{--} \label{phi--} \\
\phi^{-+} & = & G_0 \psi^{-+} \label{phi-+} \\
\phi^{++} & = & G_0 \psi^{++} \label{phi++}
\eea
and we also define
\be
\phi = \phi^{--} + \phi^{-+} + \phi^{++} = G_0 \psi = G_0 M u u. \label{phi}
\ee

We have carried out direct numerical simulations to calculate
$\psi$- and $\phi$-fields in order to investigate their
properties.

\section{NUMERICAL SIMULATIONS}

\begin{figure}
\centerline{\psfig{figure=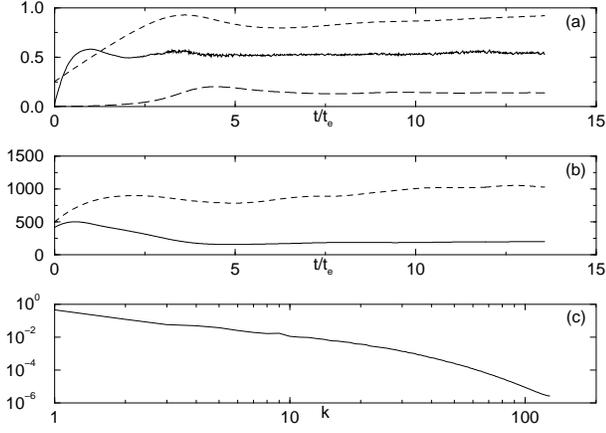,width=8truecm}}
\caption[]{\small\sf
         {\bf Simulation output.}\\
         (a) skewness (\thinline), total energy (\thindashed) and
         dissipation rate (\thinlongdashed);
         (b) microscale Reynolds number (\thinline) and
         integral scale Reynolds number (\thindashed);
         (c) evolved energy spectrum.
         ($t_e$ is the eddy turnover time)
        }
\label{simdata fig}
\end{figure}

We started from an existing, well validated code for the direct numerical
simulation (DNS) of turbulence, constructed at the University of 
Edinburgh and running on the Cray T3D administered by the
Edinburgh Parallel Computing Centre.  In Figure \ref{simdata fig}
we have plotted a number of the standard DNS outputs generated
by our code, running on a $256^3$ grid.

For the fundamentals
of direct numerical simulations, the reader is directed to
the pioneering work of Orszag (1969 and 1971).  Time
integration is performed by way of a second-order Runge-Kutta
scheme and partial dealiasing is achieved through application
of a random-shifting method (see, for example, Rogallo, 1981).
At each time-step, the nonlinear term, $\psi$, is calculated by a
pseudospectral method involving a number of fast Fourier
transforms.
$\psi^{--}$ and $\psi^{++}$ may be computed by carrying out the same
procedure having first zeroed the $u^+$ and $u^-$ fields
respectively while  $\psi^{-+}$ may be calculated simply by
subtracting $\psi^{--}$ and $\psi^{++}$ from the total nonlinear
term.

\subsection{Forcing}

Stationarity is obtained by use of a deterministic forcing
term given by,
\be
f_{\alpha} (\Vec{k}, t) = \left\{ \begin{array}{l l}
\varepsilon_0 u_\alpha(\Vec{k}, t) / [2E_f(t)] & \mbox{if $0<k<k_f$,} \\
0 & \mbox{otherwise,}
\end{array} \right.
\ee
where $\varepsilon_0$ is the {\sl desired} mean dissipation rate
(supplied as an input parameter to the simulation), and
\be
E_f(t) = \int_0^{k_f} E(k,t) \mbox{d}k
\ee
with $E(k,t)$ defined as the energy spectrum.
$k_f$ is chosen to be $1.5$ so that the forcing is applied to only
the first shell of wavenumbers.  With this forcing, we have observed
over many simulations that
after a sufficient number of time steps the velocity field reaches
a statistically stationary form, as desired.

\subsection{Computing the $\phi$-fields}

The evolution of $\phi$ and its partitions, as
defined in equations (\ref{phi--})--(\ref{phi}), is a costly
exercise as it must be
performed in parallel with the evolution of the velocity field.
We use a simple trapezoidal method to carry out the necessary
time integrals, but the need to calculate each of the partitions
of $\psi$ at each time step
leads to a code that is roughly three times as computationally
expensive as a straightforward DNS.

Further problems arise
out of the necessity to choose suitable initial
$\phi$-fields with which to begin the computation.  Here,
we have chosen to begin by zeroing each of the
$\phi$-fields for all values of $\Vec{k}$, in the
expectation that after a sufficient number of time steps the
initial conditions will have been forgotten.  To this
end, we run the DNS code until the four $\phi$-fields
have reached a statistically stationary state,
and assume that this
indicates convergence to their true values.
\begin{figure}
\centerline{\psfig{figure=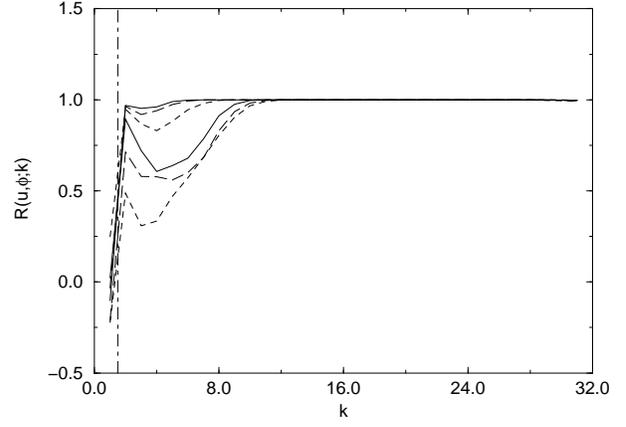,width=8truecm}}
\caption[]{\small\sf
         Correlation between $u$ and $\phi$ at
         $t=0.5t_e$ (\thindashed);
         $t=1t_e$ (\thinlongdashed);
         $t=2t_e$ (\thinline);
         $t=4t_e$ (\thickdashed);
         $t=6t_e$ (\thicklongdashed);
         $t=8t_e$ (\thickline) where $t_e$ is the
         evolved eddy turnover time.
         The dot-dashed line indicates
         the position of $k_f$.
        }
\label{u:phi fig}
\end{figure}
Further evidence may be provided by keeping in mind
the fact that our forcing is only being applied to a single
shell in wavenumber space, so that we have
\be
u = \phi \mbox{ for } k_f < k < k_0. \label{u=phi}
\ee
\begin{figure}[ht]
\centerline{\psfig{figure=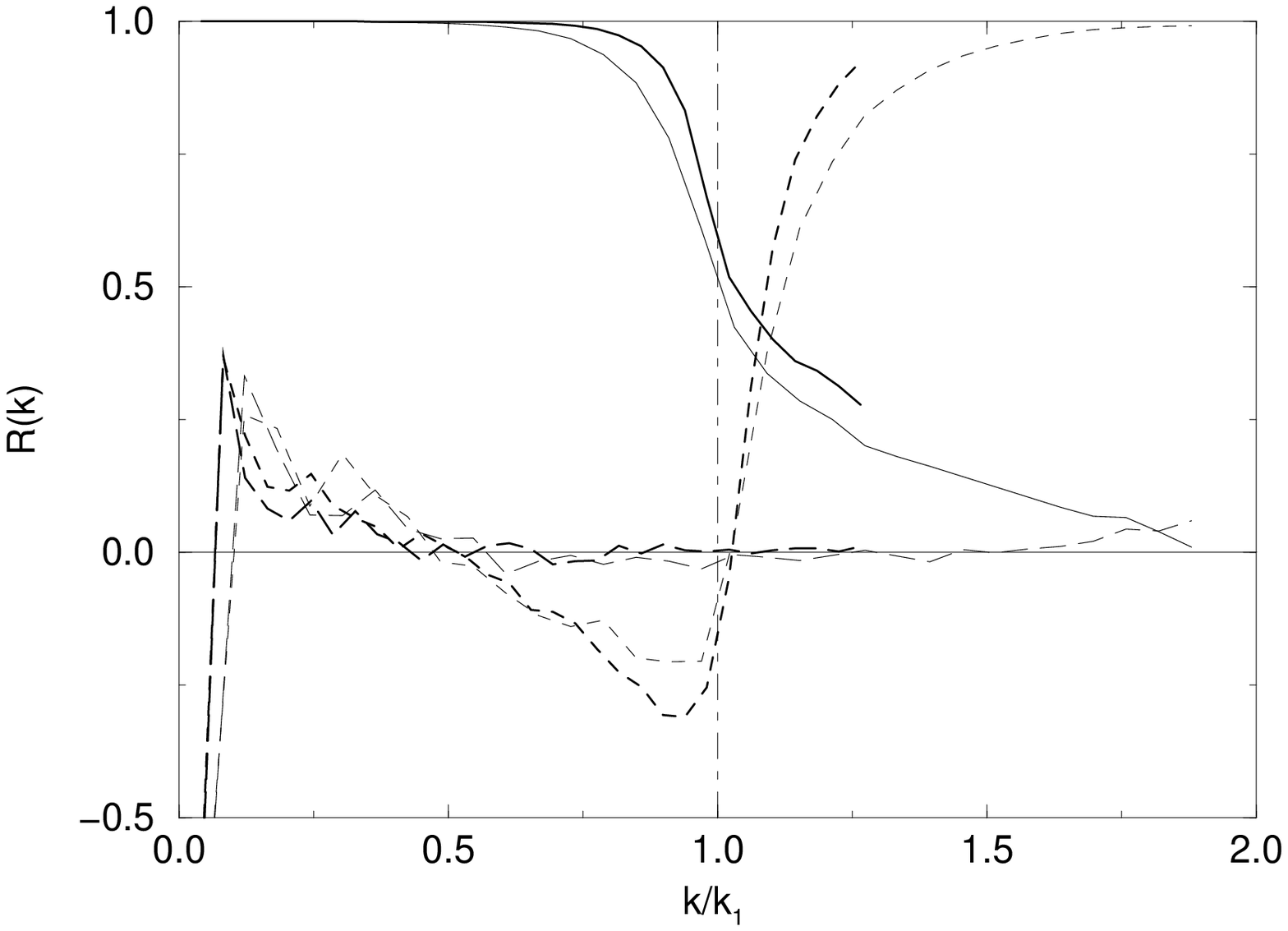,width=8truecm}}
\caption[]{\small\sf
         {\bf Integrated partitions.}\\
         $R(\phi,\phi^{--};k)$ with
         $k_1=24.5$ (\thickline) and $k_1=16.5$ (\thinline);
         $R(\phi,\phi^{-+};k)$ with
         $k_1=24.5$ (\thickdashed) and $k_1=16.5$ (\thindashed);
         $R(\phi,\phi^{++};k)$ with
         $k_1=24.5$ (\thicklongdashed) and $k_1=16.5$ (\thinlongdashed).
         The dot-dashed line indicates $k=k_1$.
        }
\label{phicorrs1 fig}
\end{figure}
\begin{figure}[ht]
\centerline{\psfig{figure=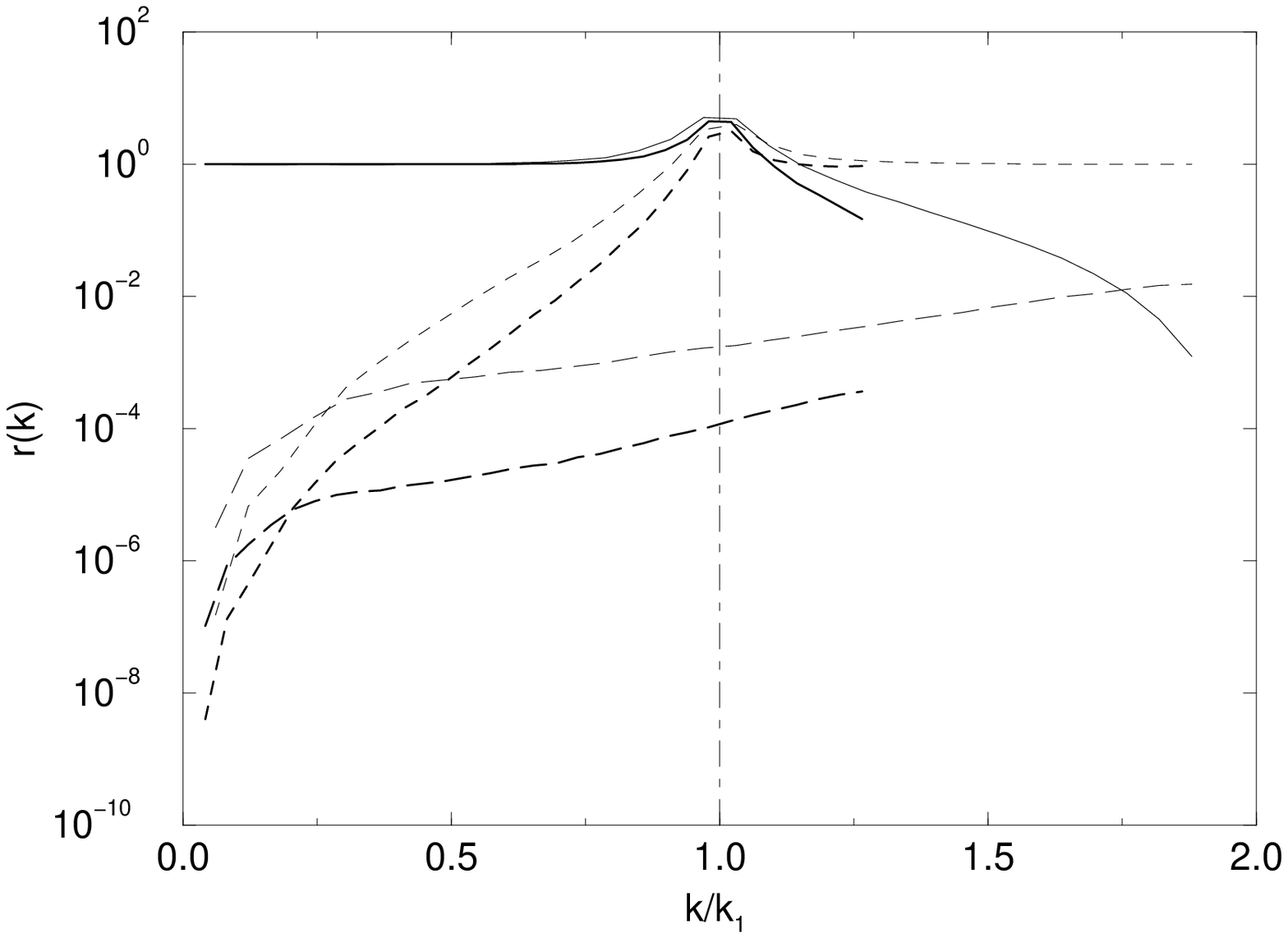,width=8truecm}}
\caption[]{\small\sf
         {\bf Integrated partitions.}\\
         $r(\phi,\phi^{--};k)$ with
         $k_1=24.5$ (\thickline) and $k_1=16.5$ (\thinline);
         $r(\phi,\phi^{-+};k)$ with
         $k_1=24.5$ (\thickdashed) and $k_1=16.5$ (\thindashed);
         $r(\phi,\phi^{++};k)$ with
         $k_1=24.5$ (\thicklongdashed) and $k_1=16.5$ (\thinlongdashed).
         The dot-dashed line indicates $k=k_1$.
        }
\label{phicorrs2 fig}
\end{figure}
\begin{figure}[ht]
\centerline{\psfig{figure=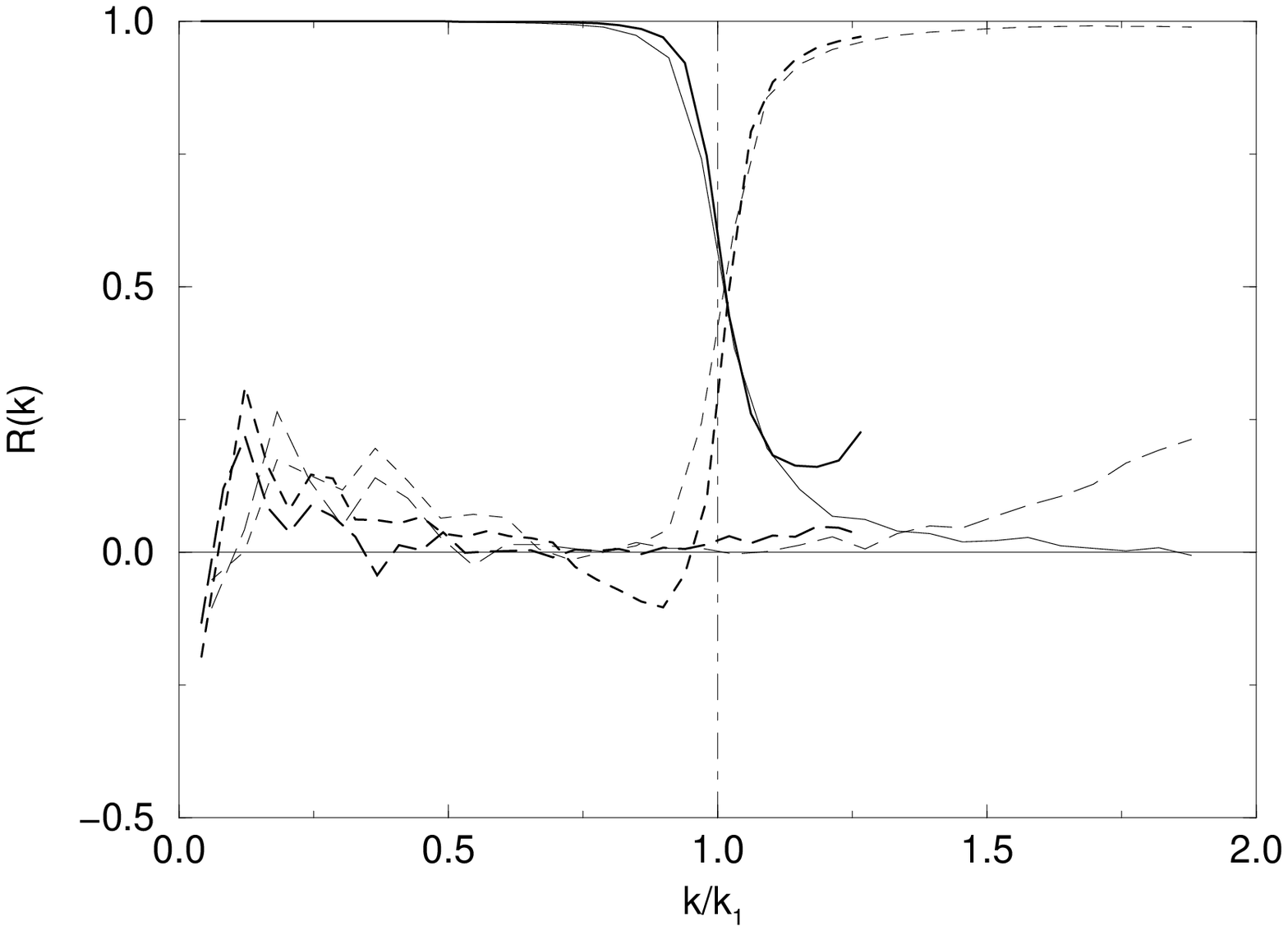,width=8truecm}}
\caption[]{\small\sf
         {\bf Instantaneous partitions.}\\
         $R(\psi,\psi^{--};k)$ with
         $k_1=24.5$ (\thickline) and $k_1=16.5$ (\thinline);
         $R(\psi,\psi^{-+};k)$ with
         $k_1=24.5$ (\thickdashed) and $k_1=16.5$ (\thindashed);
         $R(\psi,\psi^{++};k)$ with
         $k_1=24.5$ (\thicklongdashed) and $k_1=16.5$ (\thinlongdashed).
         The dot-dashed line indicates $k=k_1$.
        }
\label{psicorrs1 fig}
\end{figure}
\begin{figure}[ht]
\centerline{\psfig{figure=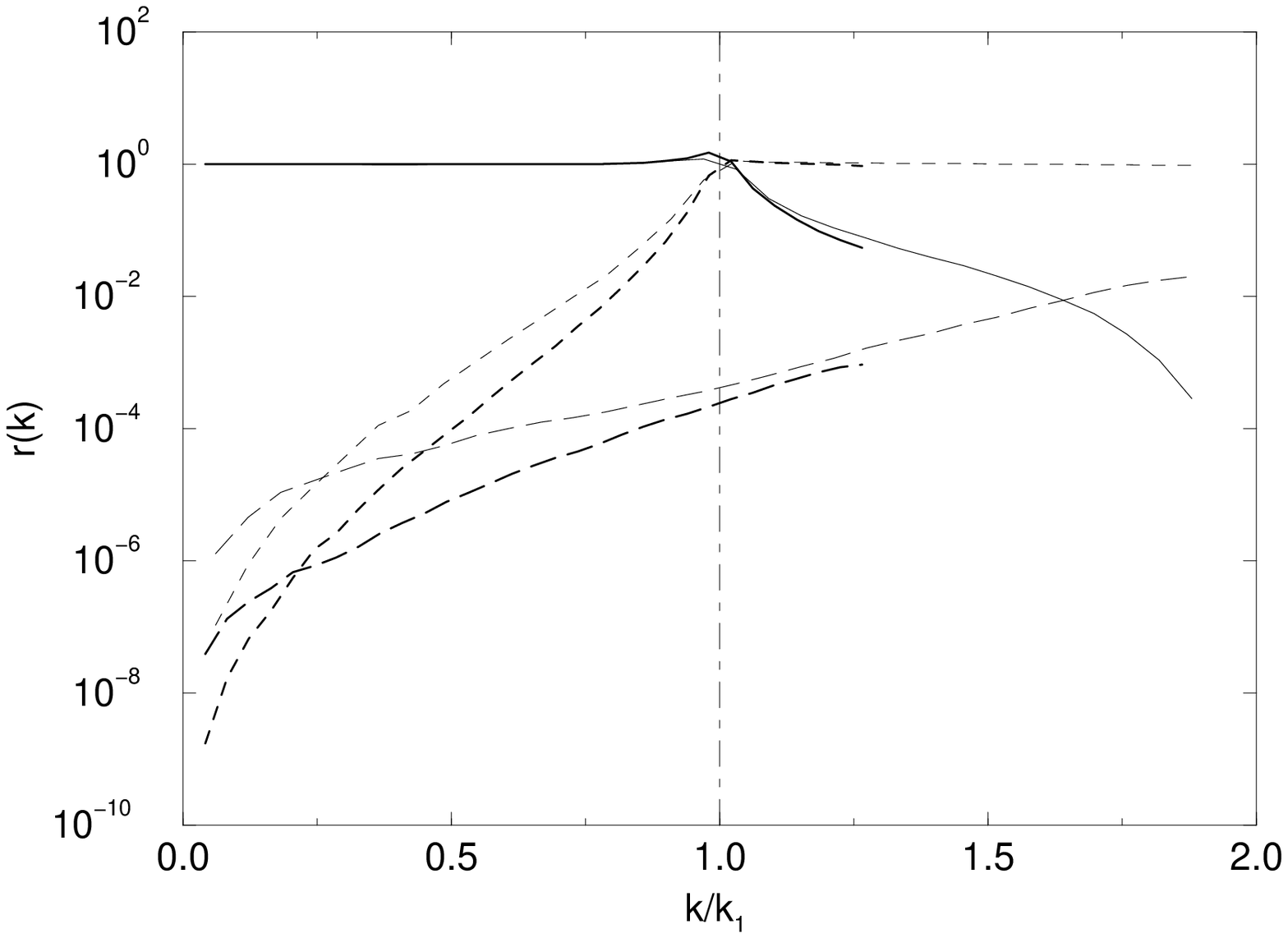,width=8truecm}}
\caption[]{\small\sf
         {\bf Instantaneous partitions.}\\
         $r(\psi,\psi^{--};k)$ with
         $k_1=24.5$ (\thickline) and $k_1=16.5$ (\thinline);
         $r(\psi,\psi^{-+};k)$ with
         $k_1=24.5$ (\thickdashed) and $k_1=16.5$ (\thindashed);
         $r(\psi,\psi^{++};k)$ with
         $k_1=24.5$ (\thicklongdashed) and $k_1=16.5$ (\thinlongdashed).
         The dot-dashed line indicates $k=k_1$.
        }
\label{psicorrs2 fig}
\end{figure}
We note that $\langle |\Vec{u}|^2 \rangle \approx \langle |\Vec{\phi}|^2
\rangle$
from our evolved data but carry out an additional test by computing
the correlation between the two fields.  We define the
general correlation between two fields, $a$ and $b$, by
\be
R(a,b;k) = \frac{\langle a_\alpha(\Vec{k}) b_\alpha(-\Vec{k}) \rangle}
{\langle |\Vec{a}(\Vec{k})|^2 \rangle^{1/2}
\langle |\Vec{b}(\Vec{k})|^2 \rangle^{1/2}} \label{corr}
\ee
and plot $R(u,\phi;k)$ in Figure \ref{u:phi fig} for
six different time steps.
We see that by the final time step, the level of correlation
is excellent for $k>5$ and
good for $k>1$.  The increasing quality of correlation with
increasing $k$ is to be expected as a consequence of the
fact that higher wavenumbers evolve at a greater rate than
lower wavenumbers --- something which is borne out by looking
at the correlations computed at earlier time steps.
The deviation in the first shell is to
be expected as this is outside the valid range of
equation (\ref{u=phi}).

Finally we note that in order to compute $\phi$-fields for
different cutoff wavenumbers, $k_1$, we must reperform
the entire DNS from initial conditions up to the fully
evolved state.

\subsection{Experimental Details}

We have computed $\psi$-fields for a number of different
cutoff wavenumbers at resolutions of $64^3$ and $256^3$
grid points.  The high cost of calculating the
$\phi$-fields means that for these we have been restricted
to a resolution of $64^3$ grid points and only two
cutoff wavenumbers.

Our $64^3$ simulation achieved a microscale Reynolds number
of $R_\lambda \approx 70$ while our $256^3$ simulation
reached $R_\lambda \approx 190$.

\section{RESULTS}

\subsection{Low Reynolds Number}

The results given in this section correspond to our
$64^3$ simulation.
We begin by presenting results for the $\phi$-fields with
cutoff wavenumbers $k_1=16.5$ and $k_1=24.5$.  Throughout
this work, the cutoff wavenumbers are chosen to be half-integers
so that they lie between two distinct shells.  For each
data set, we compute the correlation between $\phi$ and
each of its partitions using equation (\ref{corr}) and also
a measure of their relative magnitudes, $r(k)$, given by
\be
r(a,b;k) = \frac{\langle |\Vec{b}(\Vec{k})|^2 \rangle}
{\langle |\Vec{a}(\Vec{k})|^2 \rangle}. \label{r}
\ee

Results are plotted in Figures \ref{phicorrs1 fig}
and \ref{phicorrs2 fig}, with
both functions plotted against $k/k_1$.  We first note
that, once scaled in this way, the exact choice of
$k_1$ seems to have little effect on the shape of the
graphs.
\comment{
As before, we plot the first wavenumber shell,
$k=1$, for completeness only --- the presence of forcing
in this region makes it impossible to draw any sensible
conclusions.}
\begin{figure}
\centerline{\psfig{figure=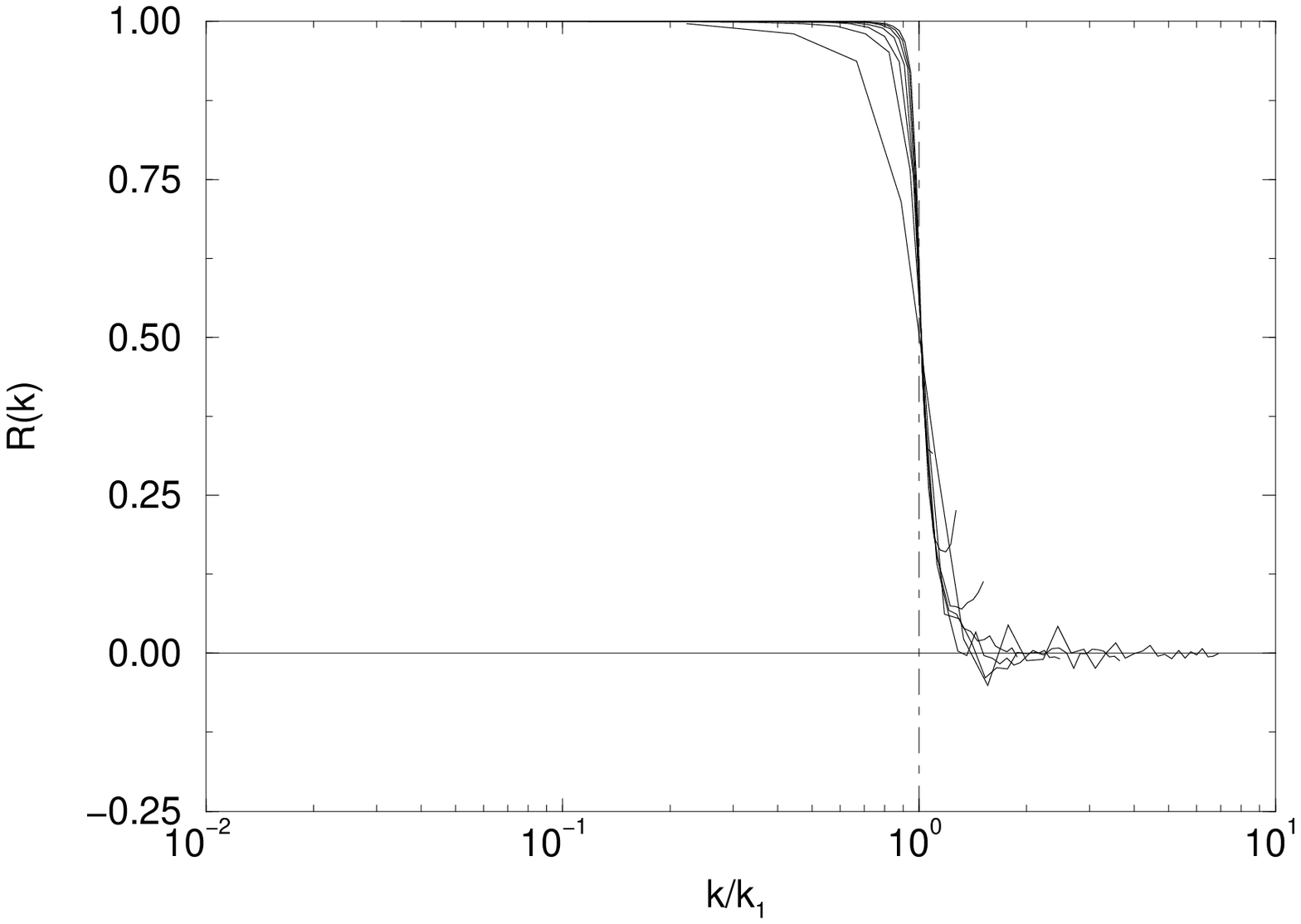,width=8truecm}}
\caption[]{\small\sf
         {\bf Low Reynolds number ($R_\lambda \approx 70$)}\\
         $R(\psi,\psi^{--};k)$ for cutoff wavenumbers
         $k_1=4.5$, $8.5$, $12.5$, $16.5$, $20.5$, $24.5$, $28.5$ with
	 $k_0=32$.  The dot-dashed line indicates $k=k_1$.
        }
\label{psimmcorrs fig}
\end{figure}
\begin{figure}
\centerline{\psfig{figure=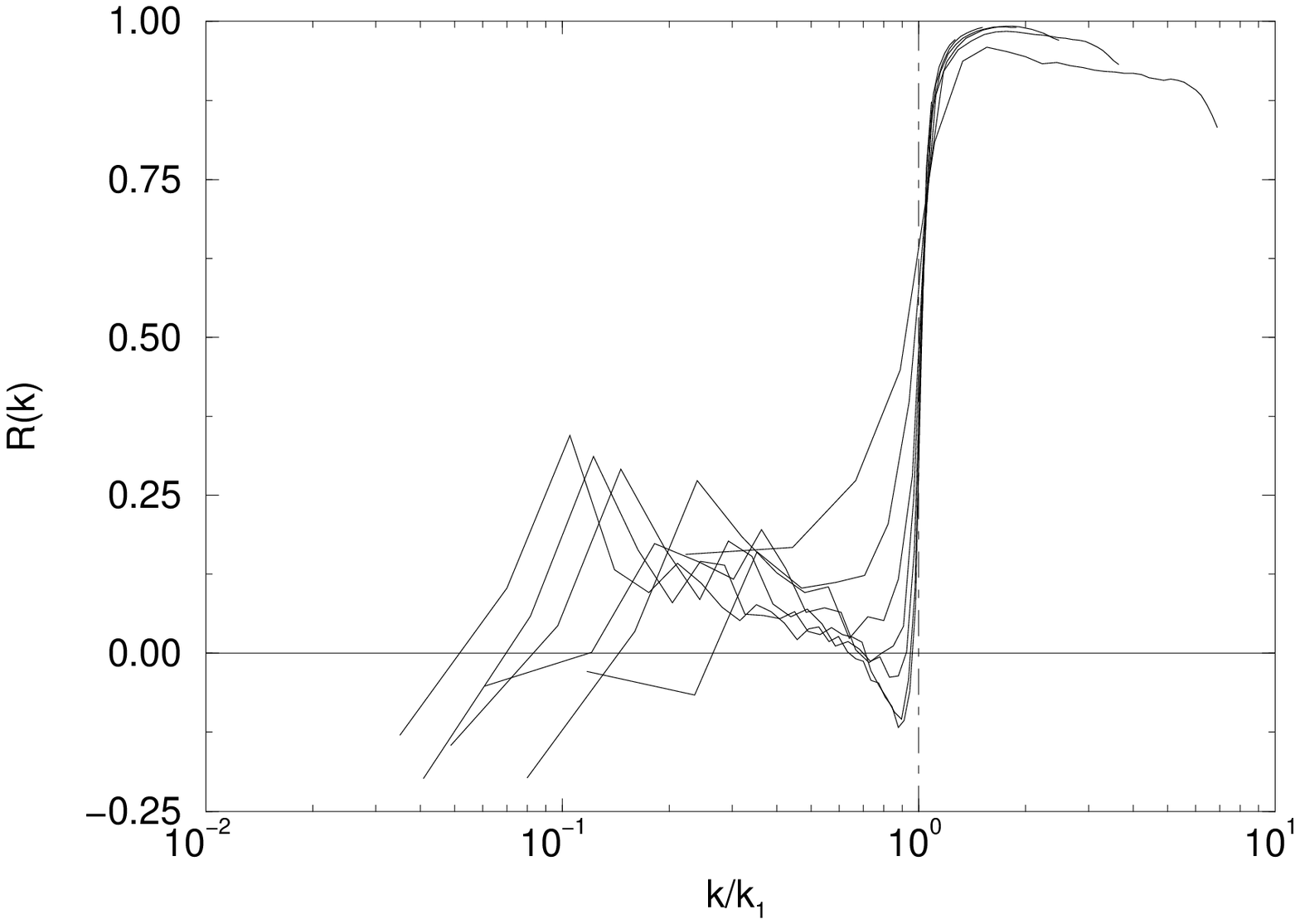,width=8truecm}}
\caption[]{\small\sf
         {\bf Low Reynolds number ($R_\lambda \approx 70$)}\\
         $R(\psi,\psi^{-+};k)$ for cutoff wavenumbers
         $k_1=4.5$, $8.5$, $12.5$, $16.5$, $20.5$, $24.5$, $28.5$ with
	 $k_0=32$.  The dot-dashed line indicates $k=k_1$.
        }
\label{psimpcorrs fig}
\end{figure}
\begin{figure}
\centerline{\psfig{figure=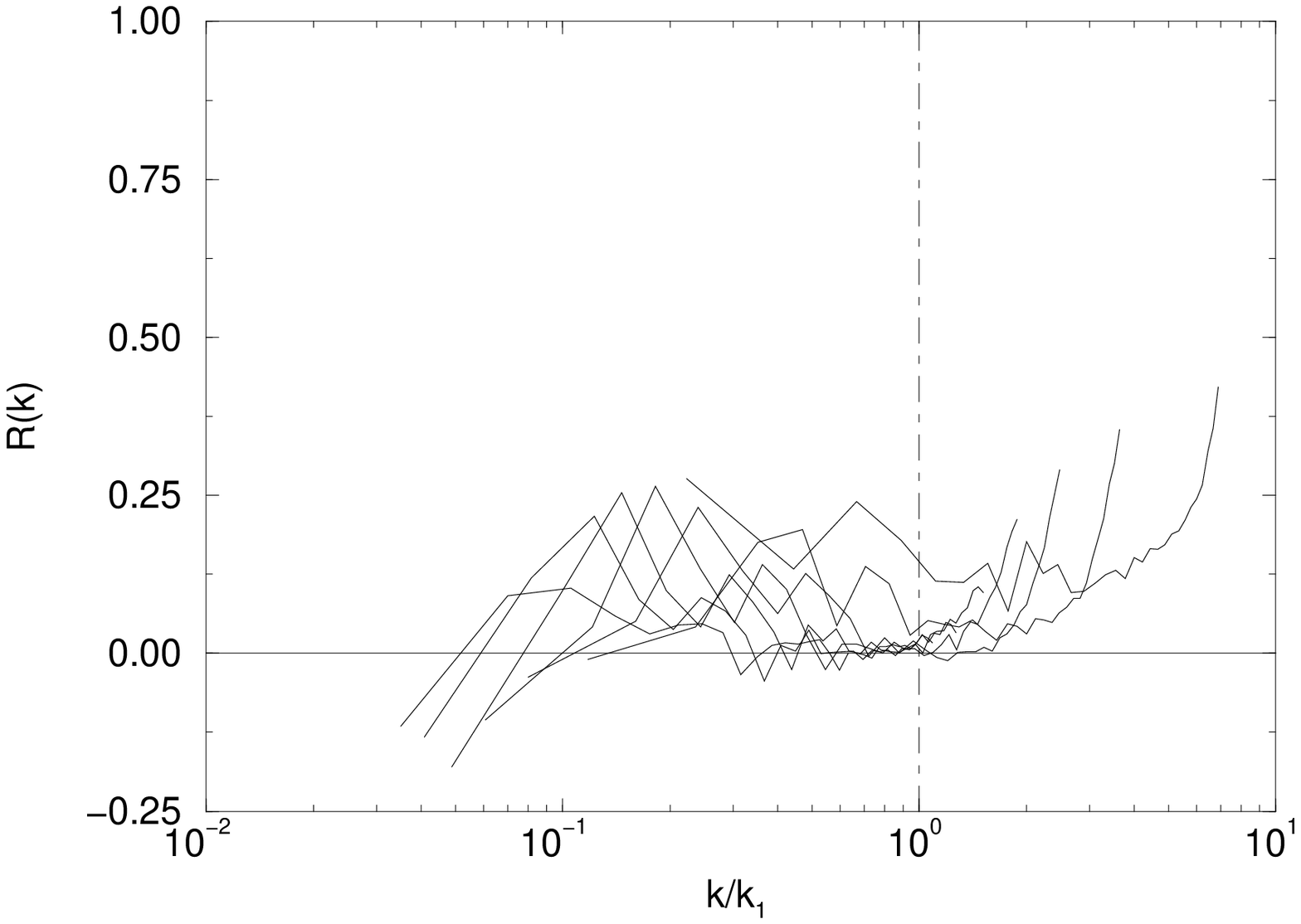,width=8truecm}}
\caption[]{\small\sf
         {\bf Low Reynolds number ($R_\lambda \approx 70$)}\\
         $R(\psi,\psi^{++};k)$ for cutoff wavenumbers
         $k_1=4.5$, $8.5$, $12.5$, $16.5$, $20.5$, $24.5$, $28.5$ with
	 $k_0=32$.  The dot-dashed line indicates $k=k_1$.
        }
\label{psippcorrs fig}
\end{figure}
\begin{figure}
\centerline{\psfig{figure=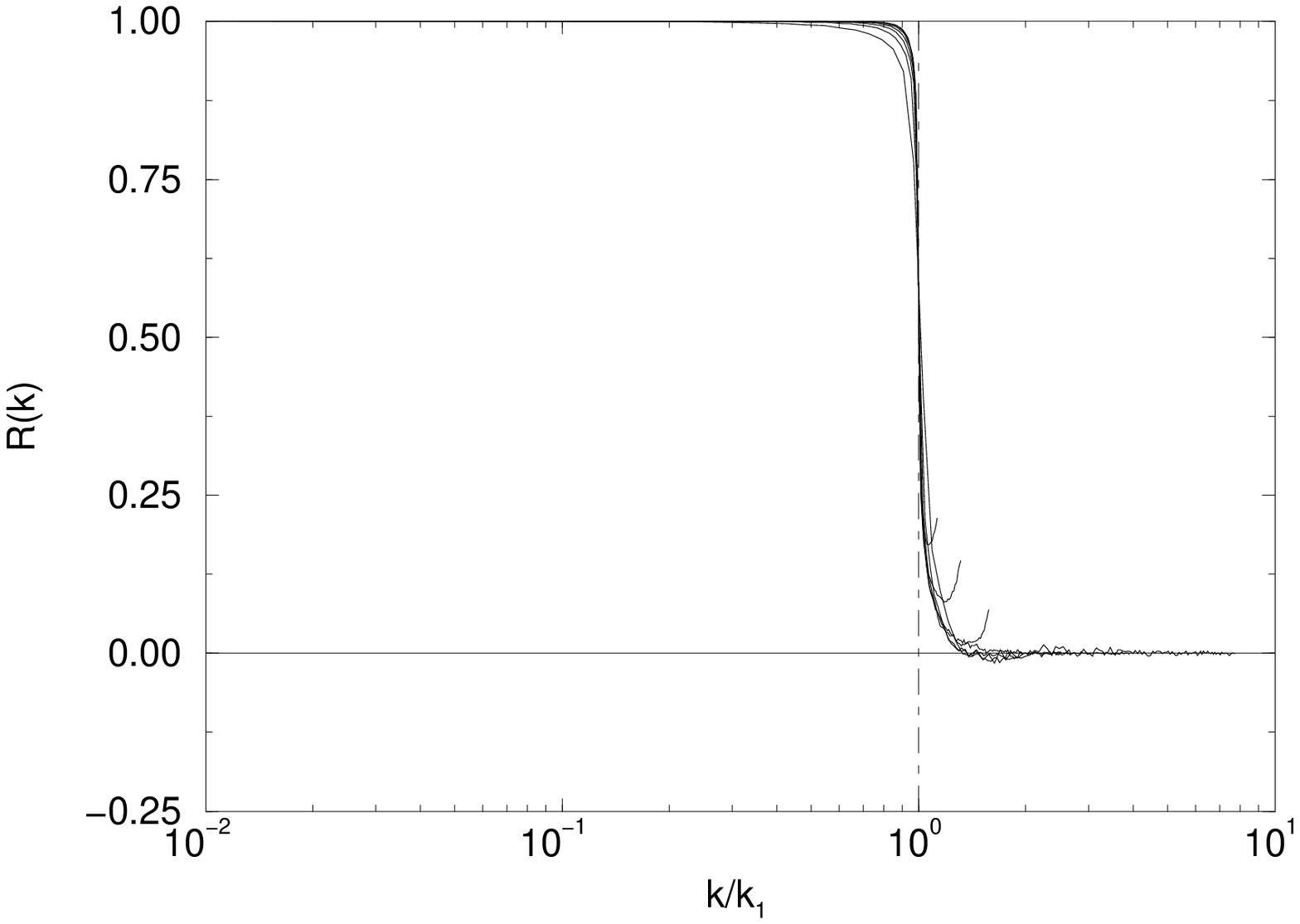,width=8truecm}}
\caption[]{\small\sf
         {\bf High Reynolds number ($R_\lambda \approx 190$)}\\
         $R(\psi,\psi^{--};k)$ for cutoff wavenumbers
         $k_1=16.5$, $32.5$, $48.5$, $64.5$, $80.5$, $96.5$, $112.5$ with
	 $k_0=128$.  The dot-dashed line indicates $k=k_1$.
        }
\label{psimmcorrs256 fig}
\end{figure}
\begin{figure}
\centerline{\psfig{figure=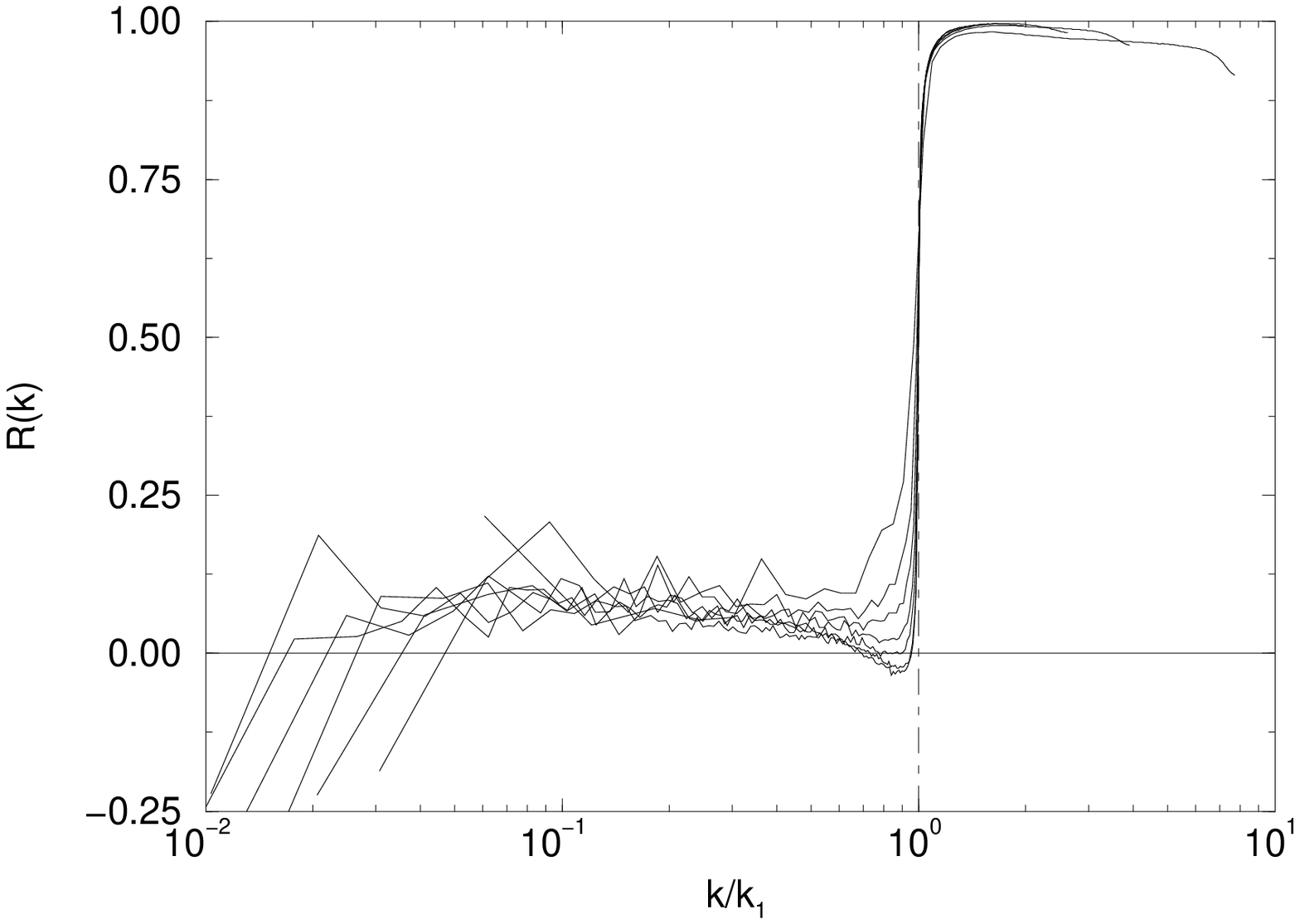,width=8truecm}}
\caption[]{\small\sf
         {\bf High Reynolds number ($R_\lambda \approx 190$)}\\
         $R(\psi,\psi^{-+};k)$ for cutoff wavenumbers
         $k_1=16.5$, $32.5$, $48.5$, $64.5$, $80.5$, $96.5$, $112.5$ with
	 $k_0=128$.  The dot-dashed line indicates $k=k_1$.
        }
\label{psimpcorrs256 fig}
\end{figure}
\begin{figure}
\centerline{\psfig{figure=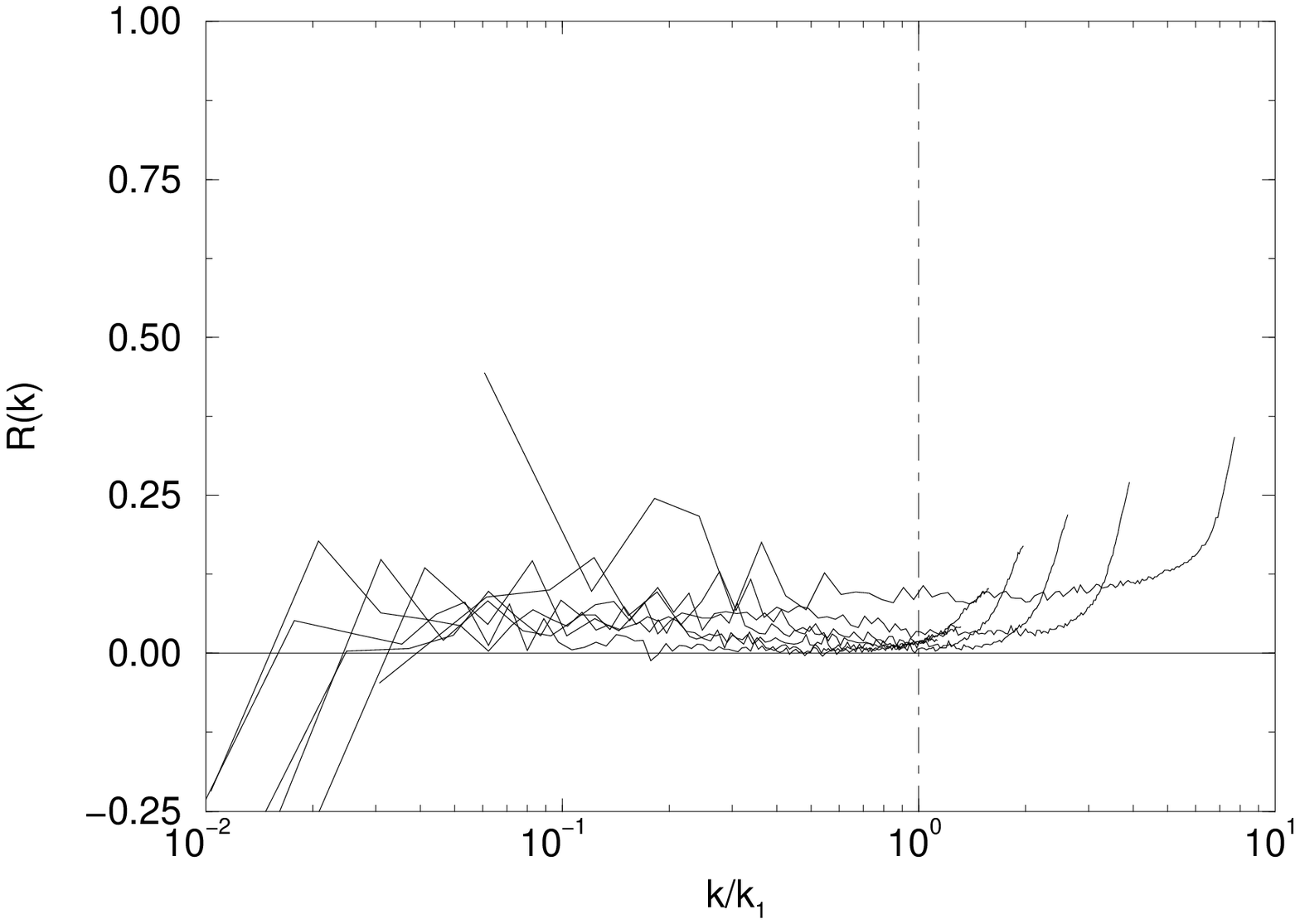,width=8truecm}}
\caption[]{\small\sf
         {\bf High Reynolds number ($R_\lambda \approx 190$)}\\
         $R(\psi,\psi^{++};k)$ for cutoff wavenumbers
         $k_1=16.5$, $32.5$, $48.5$, $64.5$, $80.5$, $96.5$, $112.5$ with
	 $k_0=128$.  The dot-dashed line indicates $k=k_1$.
        }
\label{psippcorrs256 fig}
\end{figure}
We believe that the peaky behaviour observed for
$k/k_1 < 0.5$ in Figure \ref{phicorrs1 fig}
is an effect caused by the presence
of the forcing term.

The general picture which
seems to emerge from both correlation and magnitude
information, is that for $k<k_1$, the
$\phi^{--}$ partition is the dominant part of $\phi$.  For
$k>k_1$, it is $\phi^{-+}$ which is dominant while
$\phi^{++}$ is broadly insignificant for all values of
$k$.

It was at this point in our work that it became apparent that
the computational cost involved in calculating the $\phi$-fields
was too high and so attention was turned to the $\psi$-fields which
are far easier to calculate.  For these, a single velocity
field realization is enough to calculate $\psi$ and its
partitions for any cutoff wavenumber, $k_1$.
We begin by duplicating the calculations
performed on the $\phi$-fields for the same resolution grid, and
for the same cutoff wavenumbers.  Results are shown in
Figures \ref{psicorrs1 fig} and \ref{psicorrs2 fig} where we
see a very similar picture
to that presented in Figures \ref{phicorrs1 fig} and
\ref{phicorrs2 fig}.

Taking advantage of the reduction in computational effort,
we now compute $R(\psi,\psi^{--};k)$ for a number
of different cutoff wavenumbers.  The results are shown
in Figure \ref{psimmcorrs fig}.  We see first of all that
there is an excellent collapse of data for all cutoffs
considered and that there is good correlation
between $\psi$ and $\psi^{--}$ below the cutoff.  Above
the cutoff, this correlation decays rapidly away to zero.
We also note that, mathematically, $\psi^{--}(\Vec{k})=0$
for $k>2k_1$ so that the occurrence of a non-zero
correlation in this region points to the existence
of small numerical and aliasing errors.

This picture is reversed when we consider $R(\psi,\psi^{-+};k)$
as shown in Figure \ref{psimpcorrs fig}.  The collapse of
data is not so good, particularly in the low-wavenumber
region, and we see that as $k/k_1$ increases, the correlation
begins to tail away from unity.

A pattern is even more difficult to discern when we compute
$R(\psi,\psi^{++};k)$, shown in Figure \ref{psippcorrs fig}.
We see that overall, $\psi^{++}$ does not correlate well
with $\psi$, although the level of correlation increases
with $k/k_1$ as we move beyond the cutoff.

\subsection{Moderate Reynolds Number}

We now extend our work by applying the ideas outlined
in previous sections to $\psi$-field data from our $256^3$ simulations.
Figures \ref{psimmcorrs256 fig}--\ref{psippcorrs256 fig} show
correlations of $\psi$ with each of its partitions from this data.
As can be easily seen, the picture has changed very little from
our $64^3$ data, the biggest difference being that the results
seem better behaved, which is expected due to there being
more data points available for shell averaging in the region
$k<k_1$.

In the following sections, all results are generated using
the $\psi$-fields taken from our $256^3$ simulation.

\subsection{Eddy-viscosities with Sharp Cutoffs}

We begin by rewriting our Navier-Stokes equation (\ref{L0u=partition})
for the low-wavenumber modes as
\be
L_0 u_< = \psi^{--}_< + \psi^{(--)}_< + f_< \label{NSE-}
\ee
where the subscript `$<$' indicates that we are only concerned with
$k<k_1$ and where
\be
\psi^{(--)} = \psi^{-+}+\psi^{++}.
\ee
In a large
eddy simulation, wavenumbers $k>k_1$ will not be available and so
we introduce some model for $\psi^{(--)}_<$ which we will denote
$\tilde{\psi}^{(--)}_<$.
A standard form for $\tilde{\psi}^{(--)}_<$ is an eddy-viscosity model, whereby
\be
\tilde{\psi}^{(--)}_< = -\delta \nu(k) k^2 u_< \label{ev model}
\ee
for some viscosity increment, $\delta \nu(k)$.

We now consider a hypothetical large eddy simulation, based around
the idea of our wavenumber cutoffs introduced in previous sections.

\begin{figure}
\centerline{\psfig{figure=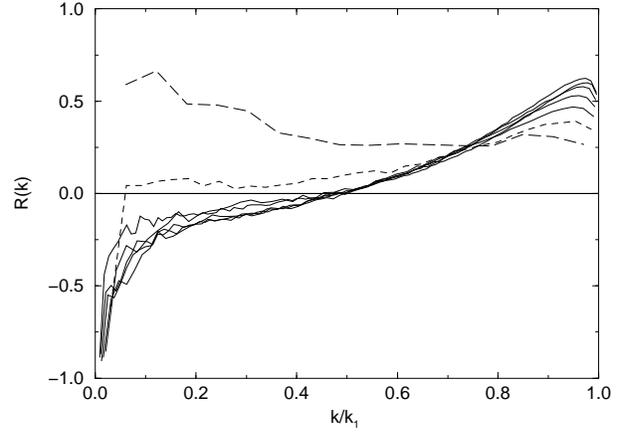,width=8truecm}}
\caption[]{\small\sf
         $R(\psi^{(--)}_<,\tilde{\psi}_<^{(--)};k)$ for cutoff wavenumbers
         $k_1=16.5$ (\thinlongdashed), $32.5$ (\thindashed),
         $48.5, 64.5, 80.5, 96.5, 112.5$ (\thinline) with
	 $k_0=128$.
        }
\label{evcorr fig}
\end{figure}

We can now compute the correlation between the exact subgrid
terms and the model, $R(\psi^{(--)}_<,\tilde{\psi}_<^{(--)};k)$.
Because of $\tilde{\psi}_<^{(--)}$'s relationship with
$u$ --- and assuming only that $\delta\nu(k)$ is positive for all
values of $k$ --- this is equal to
$R(\psi^{(--)}_<,-u_<;k)$.  The results are plotted in
Figure \ref{evcorr fig}.  We see immediately that the
correlation is, in general, quite poor.  We also see what
appears to be a difference in behaviour between cutoffs
$k_1 \ge 48.5$ where there appears to be some degree
of universality and cutoffs $k_1 \le 32.5$.  In general,
however, we can state quite categorically that, for the case
of a sharp cutoff in wavenumber space, no eddy-viscosity model
(subject to the reasonable constraint $\delta\nu(k) \ge 0$)
can perfectly reproduce the missing nonlinear terms.  But,
as we shall see, our conclusion will be modified somewhat
when we consider phase and amplitude information separately.

\begin{figure}
\centerline{\psfig{figure=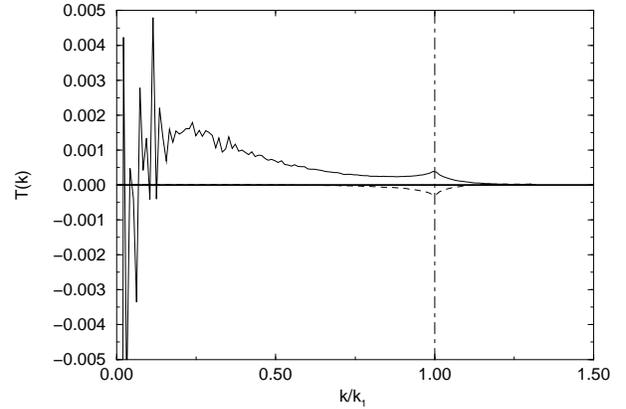,width=8truecm}}
\caption[]{\small\sf
         $T^{--}(k)$ (\thinline) and
         $T^{(--)}(k)$ (\thindashed) for cutoff wavenumber
         $k_1=96.5$ with $k_0=128$.
        }
\label{Tspec fig}
\end{figure}

\begin{figure}
\centerline{\psfig{figure=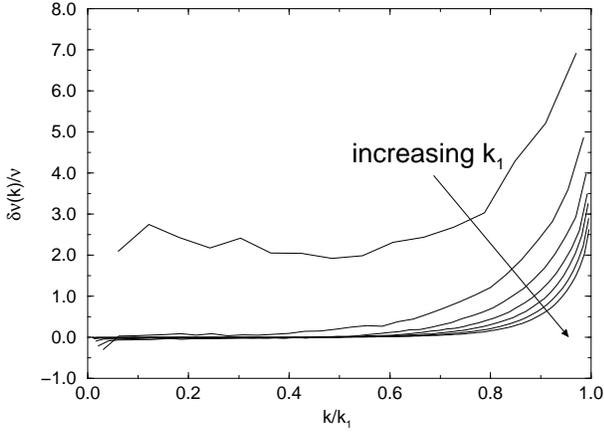,width=8truecm}}
\caption[]{\small\sf
         $\delta \nu(k)$ for cutoff wavenumbers
         $k_1=16.5, 32.5, 48.5, 64.5, 80.5, 96.5, 112.5$ with
	 $k_0=128$.
        }
\label{dnu fig}
\end{figure}

Returning to equation (\ref{NSE-}), we can generate an energy
balance equation by multiplying through by $u_<$ and averaging.
Multiplying this through by $2 \pi k^2$ then gives,
\be
\ddt E_< + 2 \nu k^2 E_< = T^{--}_< + T^{(--)}_< + W_< \label{Ebal-}
\ee
where $T^{--}_<$ describes energy transfer to and from the
low-wavenumber modes through exclusively low-wavenumber couplings while
$T^{(--)}_<$ describes energy transfer to and from the low-wavenumber
modes through coupling involving at least one high-wavenumber mode.
$W_<$ is the energy input due to the forcing.
For interest's sake, the two energy transfer functions
are plotted in Figure \ref{Tspec fig}.  We see that
$T^{--}$ has a large negative value in the first shell
(note that the $y$-axis has been truncated for this graph)
corresponding to transfer of energy away from the energy
input (forcing).  We also see that $T^{--}$ is piling up
energy at the cutoff and that this is balanced by
$T^{(--)}$ which carries it to the higher wavenumbers.

Introducing an eddy-viscosity
model as defined in equation (\ref{ev model}) will give us an
energy balance equation for LES,
\be
\ddt E_< + 2 \nu k^2 E_< + 2 \delta\nu(k) k^2 E_< = T^{--}_< + W_<
\label{Ebal LES}.
\ee
Comparing equations (\ref{Ebal-}) and (\ref{Ebal LES}) we can
easily derive a form for the eddy viscosity,
\be
\delta \nu(k) = - \frac{T^{(--)}_<}{2 k^2 E_<}.
\ee
For a large eddy simulation, this would ordinarily have to be
estimated by use of some model, as the whole point of LES is
the absence of the high-wavenumber modes necessary for the
calculation of $T^{(--)}_<$.  However, with DNS data we {\sl can}
calculate this, and the results are plotted in Figure \ref{dnu fig}.

The general form of these eddy-viscosities appears
to be in good agreement with the form
obtained theoretically by, for example, Kraichnan (1976) and we
know from previous work (see, for example, Lesieur and Rogallo,
1989) and our own LES experiments that
this particular model provides good results.  We must
now ask why this is, when it is in apparent contradiction with
the results presented in Figure \ref{evcorr fig}.

\subsection{Separating out Phase and Amplitude Effects}

In this section we look at what happens if we separate our data
into those contributions due to phase and those contributions
due to amplitude.  For each point in our field given by the
wavevector, $\Vec{k}$, we choose some unit vector perpendicular to
$\Vec{k}$ which we call $\hat{\Vec{n}}$.  We may then generate
a transverse component of the velocity field,
\be
u_T = \Vec{u} \Vec{\cdot} \hat{\Vec{n}}.
\ee
We note that $u_T$
provides us with a {\sl statistically} complete picture
of our system.  This is because, through continuity,
$u_L$, the component parallel to $\Vec{k}$ will be zero and
due to isotropy, the statistical properties of $u_T$ are
independent of the exact choice of $\hat{\Vec{n}}$.

Since we are working in Fourier space,
$u_T$ will be a complex scalar and so may be rewritten in
the form
\be
u_T = u_r e^{i u_{\theta}},
\ee
and similarly for our $\psi$-fields.
We may now consider correlations based solely on phase or
amplitude information (note that as we are now working
with {\sl scalar} fields, equation (\ref{corr}) must
be modified by the removal of the sum over components,
$\alpha$).

\begin{figure}
\centerline{\psfig{figure=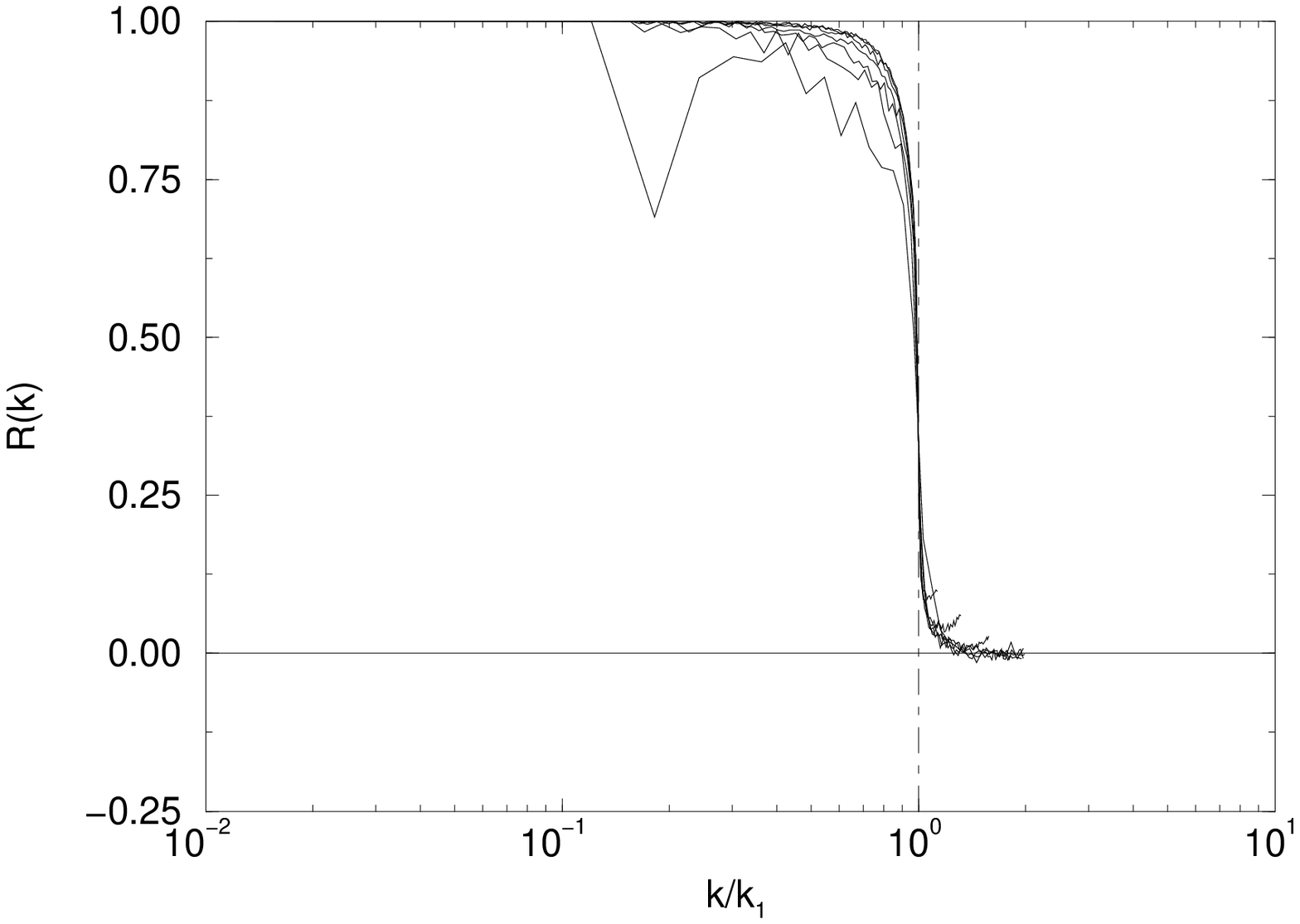,width=8truecm}}
\caption[]{\small\sf
         {\bf Phase-correlation.}\\
         $R(\psi_\theta,\psi^{--}_\theta;k)$ for cutoff wavenumbers
         $k_1=16.5$, $32.5$, $48.5$, $64.5$, $80.5$, $96.5$, $112.5$ with
	 $k_0=128$.  The dot-dashed line indicates $k=k_1$.
        }
\label{psimmcorrs_arg256 fig}
\end{figure}

\begin{figure}
\centerline{\psfig{figure=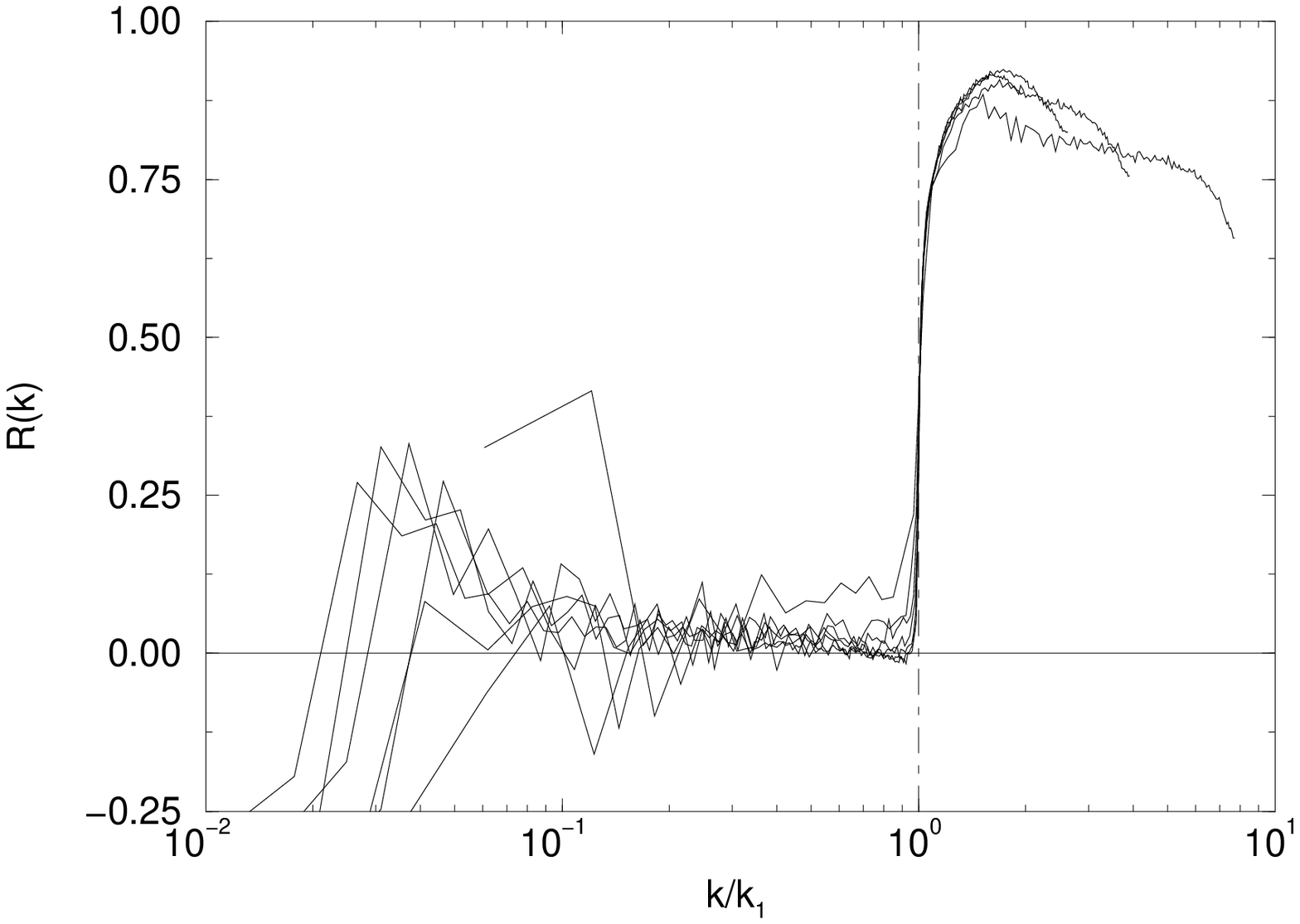,width=8truecm}}
\caption[]{\small\sf
         {\bf Phase-correlation.}\\
         $R(\psi_\theta,\psi^{-+}_\theta;k)$ for cutoff wavenumbers
         $k_1=16.5$, $32.5$, $48.5$, $64.5$, $80.5$, $96.5$, $112.5$ with
	 $k_0=128$.  The dot-dashed line indicates $k=k_1$.
        }
\label{psimpcorrs_arg256 fig}
\end{figure}

\begin{figure}
\centerline{\psfig{figure=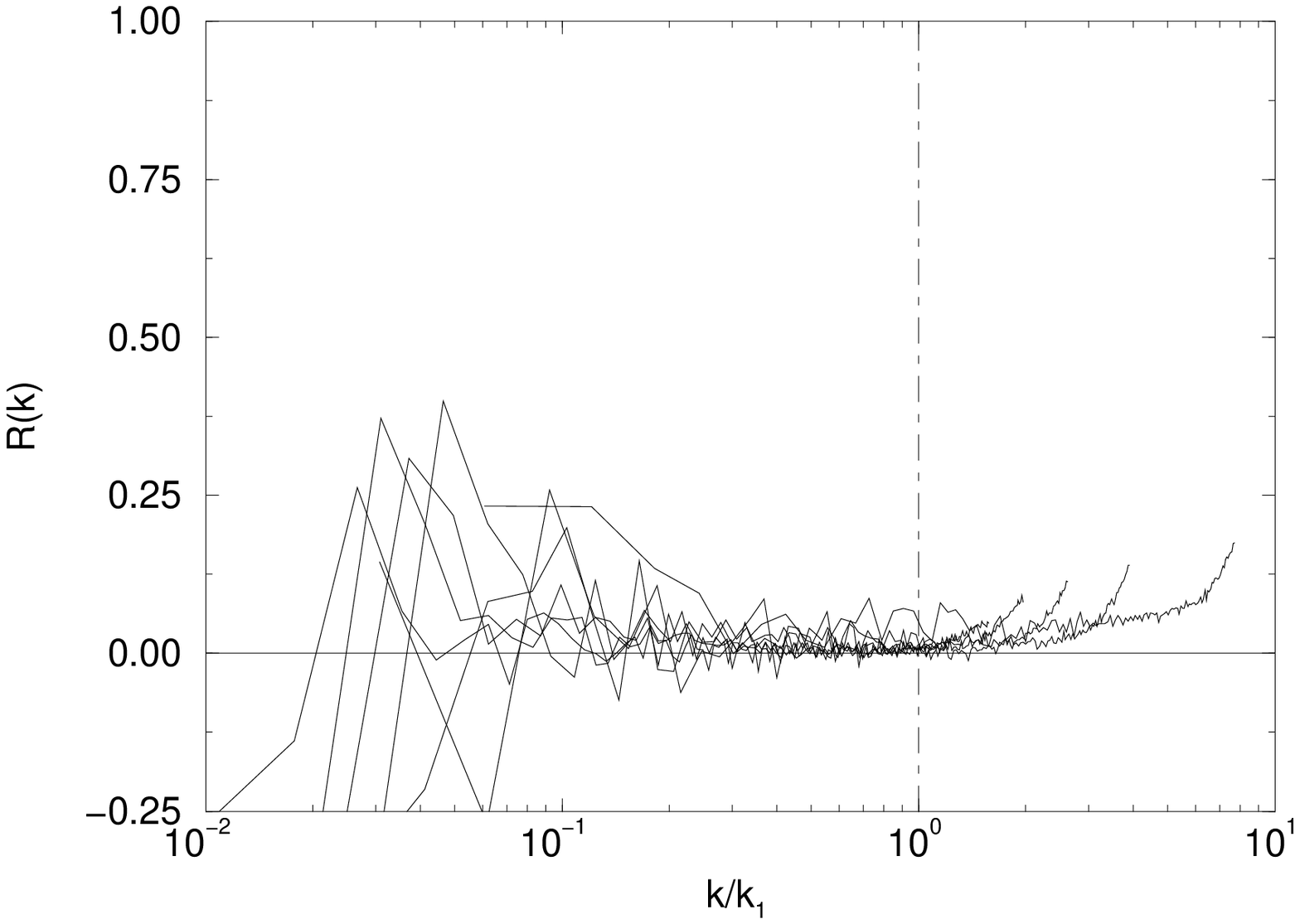,width=8truecm}}
\caption[]{\small\sf
         {\bf Phase-correlation.}\\
         $R(\psi_\theta,\psi^{++}_\theta;k)$ for cutoff wavenumbers
         $k_1=16.5$, $32.5$, $48.5$, $64.5$, $80.5$, $96.5$, $112.5$ with
	 $k_0=128$.  The dot-dashed line indicates $k=k_1$.
        }
\label{psippcorrs_arg256 fig}
\end{figure}

In Figures \ref{psimmcorrs_arg256 fig}---\ref{psippcorrs_arg256 fig} we
have plotted correlations corresponding to Figures
\ref{psimmcorrs256 fig}--\ref{psippcorrs256 fig} but concerning ourselves
only with {\sl phase} information.  We see a picture which
is largely similar to that seen when considering all parts of
the $\psi$-fields, but note that in places --- most
obviously for $k>k_1$ in the $\psi^{-+}$ data --- the correlations
are less good.

\begin{figure}
\centerline{\psfig{figure=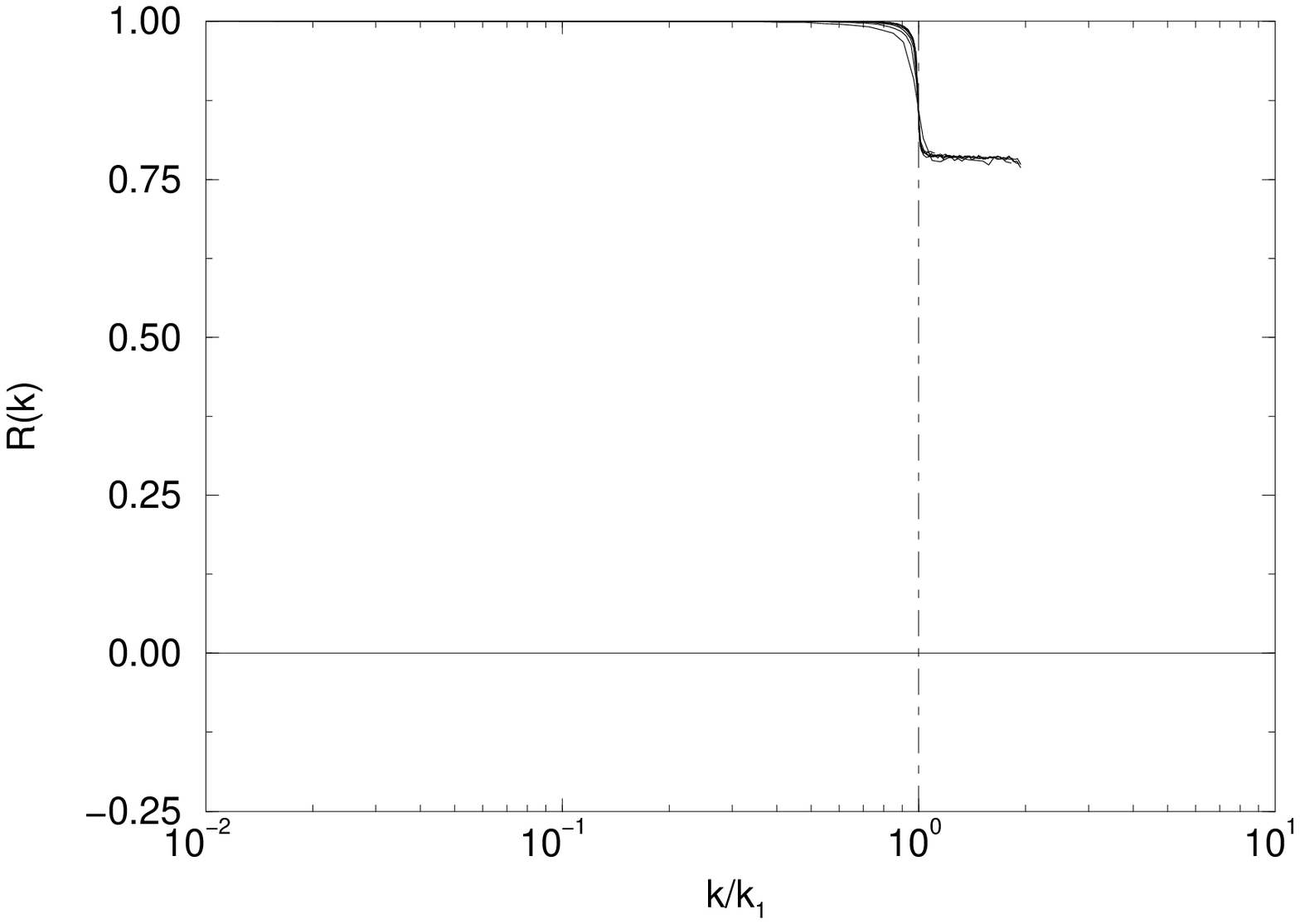,width=8truecm}}
\caption[]{\small\sf
         {\bf Amplitude-correlation.}\\
         $R(\psi_r,\psi^{--}_r;k)$ for cutoff wavenumbers
         $k_1=16.5$, $32.5$, $48.5$, $64.5$, $80.5$, $96.5$, $112.5$ with
	 $k_0=128$.  The dot-dashed line indicates $k=k_1$.
        }
\label{psimmcorrs_mag256 fig}
\end{figure}

\begin{figure}
\centerline{\psfig{figure=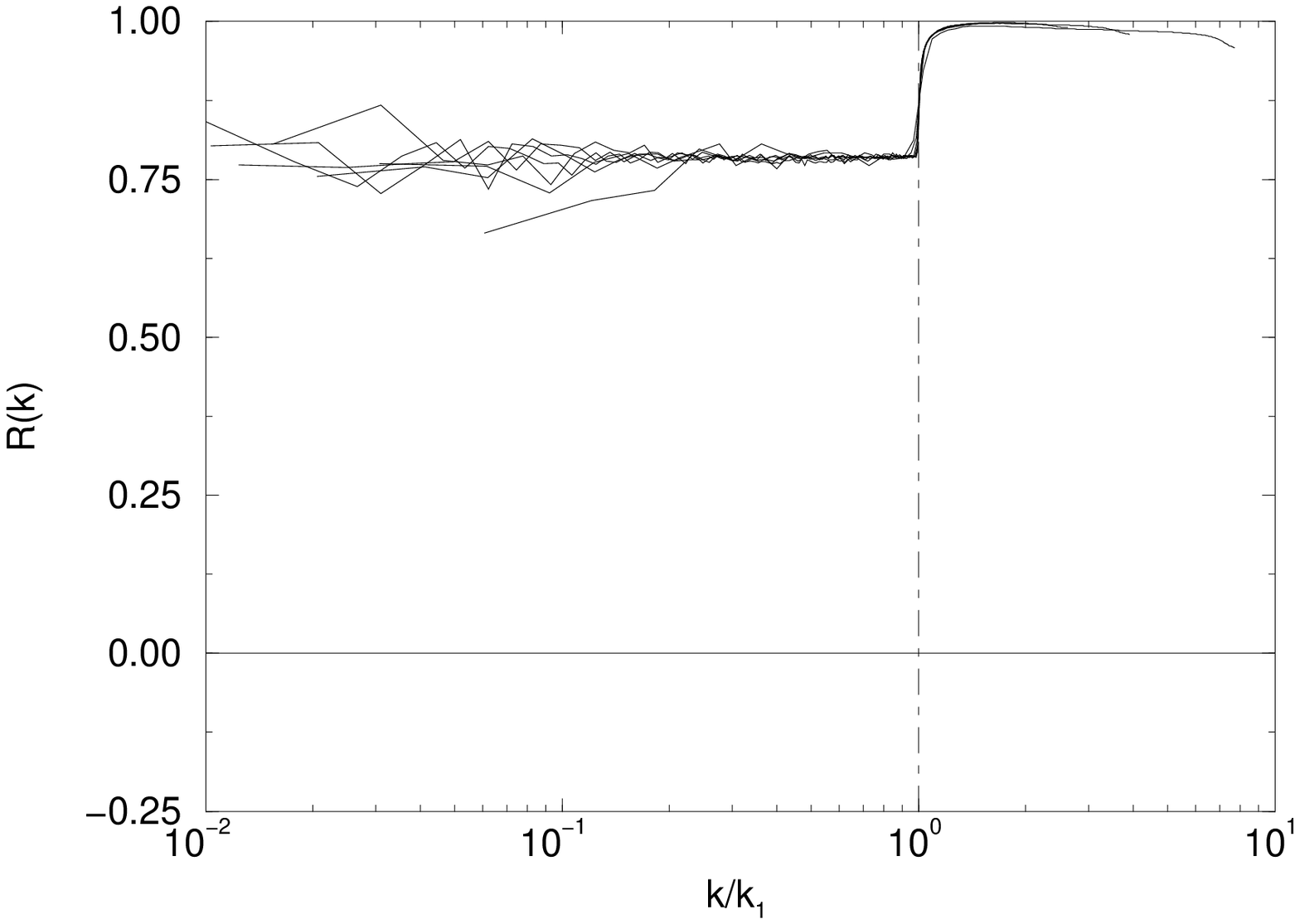,width=8truecm}}
\caption[]{\small\sf
         {\bf Amplitude-correlation.}\\
         $R(\psi_r,\psi^{-+}_r;k)$ for cutoff wavenumbers
         $k_1=16.5$, $32.5$, $48.5$, $64.5$, $80.5$, $96.5$, $112.5$ with
	 $k_0=128$.  The dot-dashed line indicates $k=k_1$.
        }
\label{psimpcorrs_mag256 fig}
\end{figure}

\begin{figure}
\centerline{\psfig{figure=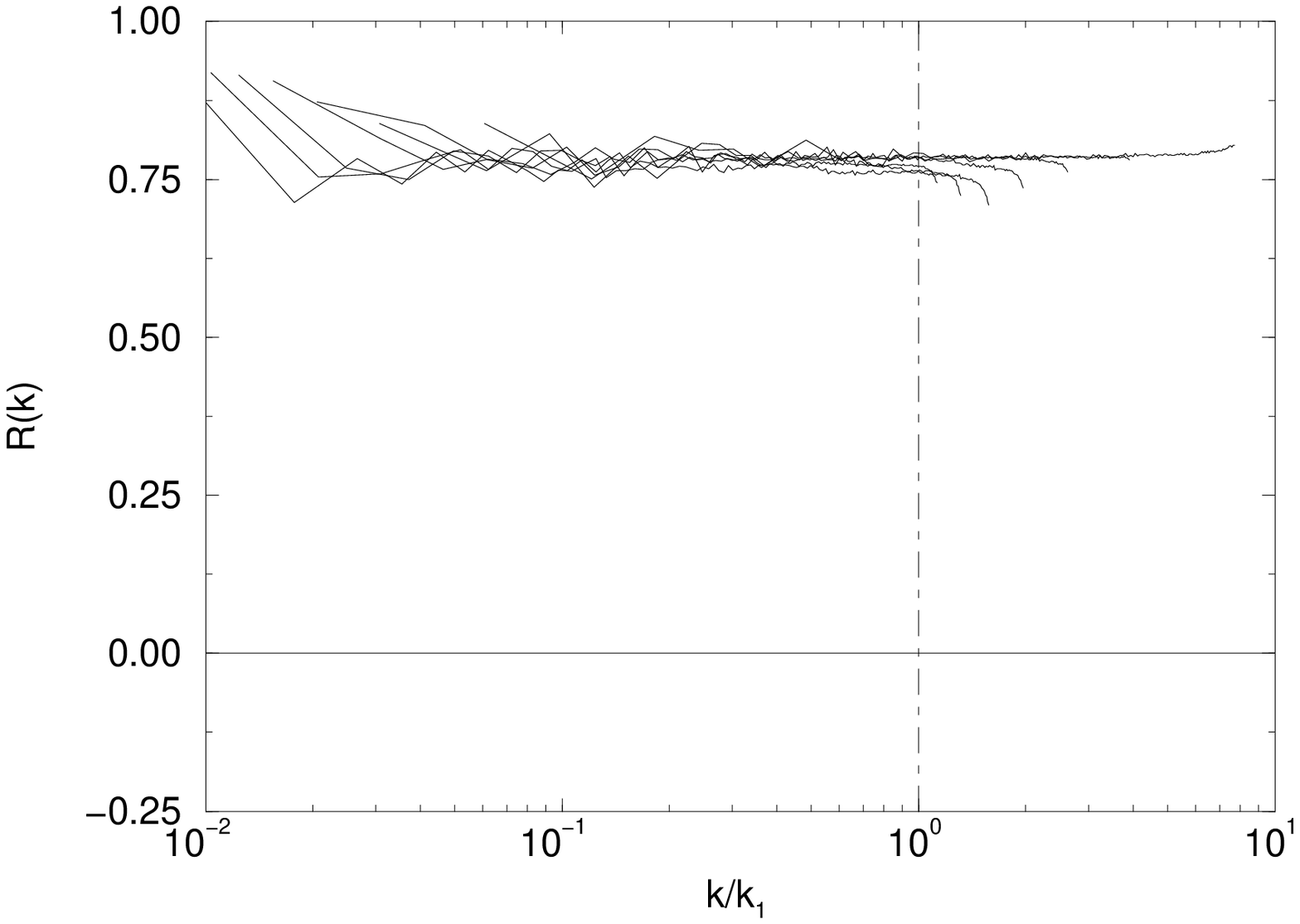,width=8truecm}}
\caption[]{\small\sf
         {\bf Amplitude-correlation.}\\
         $R(\psi_r,\psi^{++}_r;k)$ for cutoff wavenumbers
         $k_1=16.5$, $32.5$, $48.5$, $64.5$, $80.5$, $96.5$, $112.5$ with
	 $k_0=128$.  The dot-dashed line indicates $k=k_1$.
        }
\label{psippcorrs_mag256 fig}
\end{figure}

Moving on to Figures \ref{psimmcorrs_mag256 fig}---\ref{psippcorrs_mag256 fig},
where we have considered only {\sl amplitude} information,
we see a different picture.  Here, while the regions of
excellent correlation remain more or less untouched, for the
rest of the data we see that the level of correlation does not
drop below about $0.75$ and indeed seems to remain approximately
constant at some value between $0.75$ and $0.80$.

\begin{figure}
\centerline{\psfig{figure=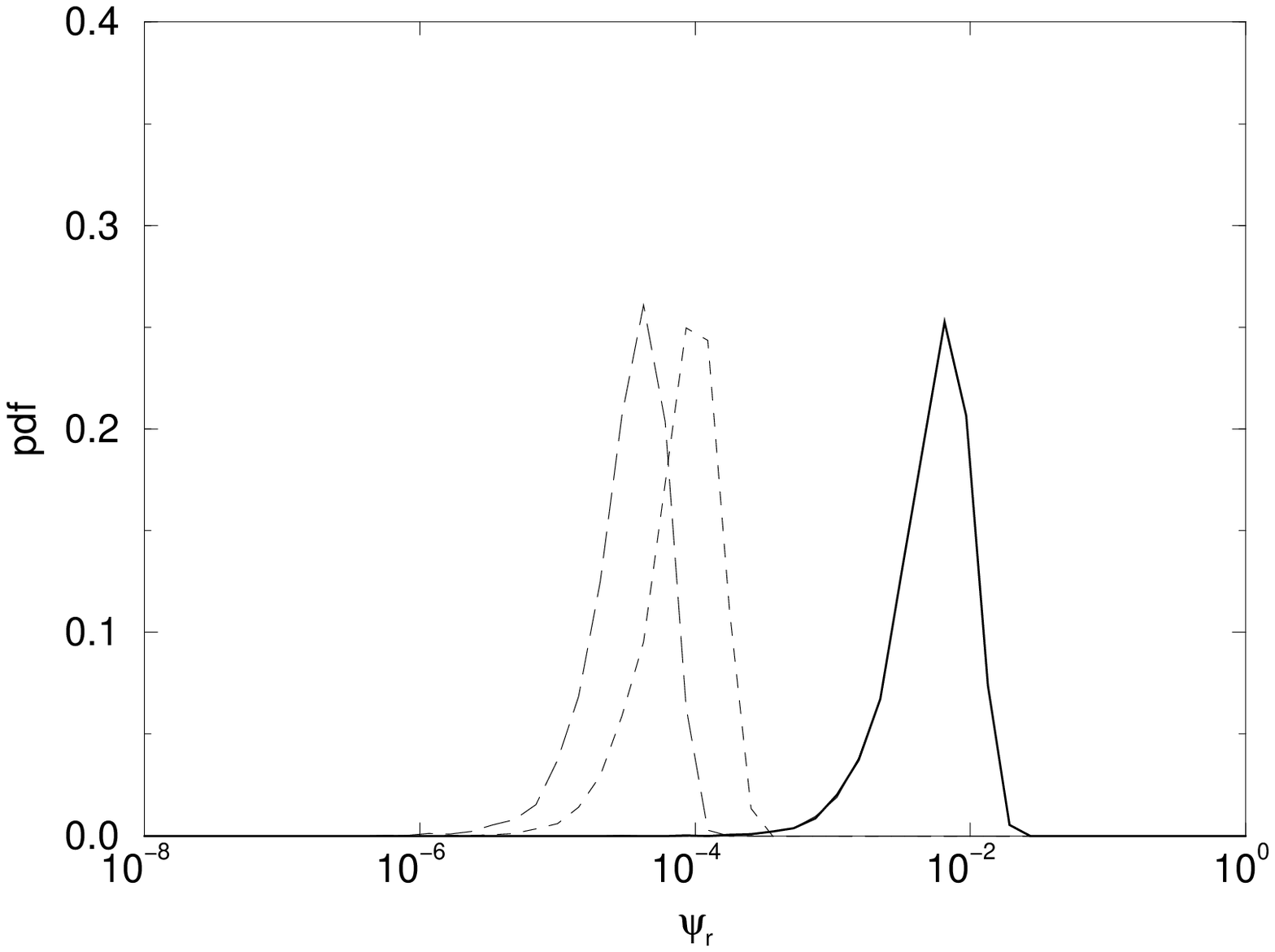,width=8truecm}}
\caption[]{\small\sf
         PDF for $\psi_r$ (\thickline);
         $\psi^{--}_r$ (\thinline);
         $\psi^{-+}_r$ (\thindashed);
         $\psi^{++}_r$ (\thinlongdashed) at $k=32$ with $k_1=64.5$.
        }
\label{PDF:Ur fig}
\end{figure}

\begin{figure}
\centerline{\psfig{figure=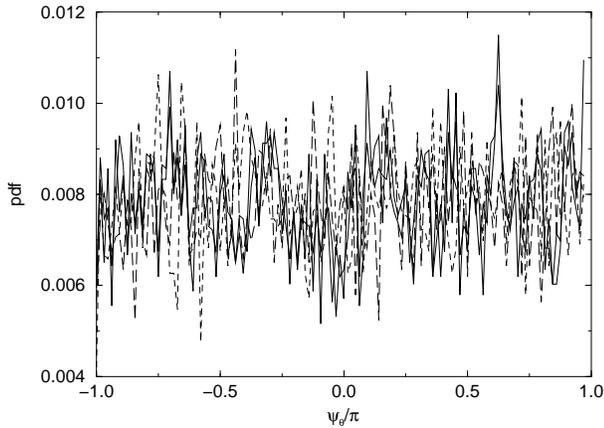,width=8truecm}}
\caption[]{\small\sf
         PDF for $\psi_\theta$ (\thickline);
         $\psi^{--}_\theta$ (\thinline);
         $\psi^{-+}_\theta$ (\thindashed);
         $\psi^{++}_\theta$ (\thinlongdashed) at $k=32$ with $k_1=64.5$.
        }
\label{PDF:Utheta fig}
\end{figure}

We can shed some light on the reasons for this by
considering the statistical nature of the $\psi$-fields.
In Figures \ref{PDF:Ur fig} and \ref{PDF:Utheta fig} we
plot probability distribution functions for the
amplitude and phase components respectively for each
of the $\psi$-fields\footnote{We note that, strictly,
these are not PDFs as our data set in this instance
is insufficiently large --- we would expect the true
PDF of the phase components to be flat, for example.  What
we have instead is a measure of the distribution of that
data which we do have.}.  We see that the possible
values of the amplitude components are localised
and similar in shape (in fact, for the chosen value of
$k<k_1$, the PDF's for $\psi_r$ and $\psi^{--}_r$ are almost
indistinguishable)
and hence amplitude correlations will be good.  On the other hand,
there is no preferred phase and hence phase correlations will
be very poor.

We now carry these ideas across to the analysis of general
eddy viscosity models outlined in a previous section.
Recall the definition of $\tilde{\psi}^{(--)}_<$ as a model
for $\psi^{(--)}_<$ --- we can now compare the phase and
amplitude components of these two fields separately.

\begin{figure}
\centerline{\psfig{figure=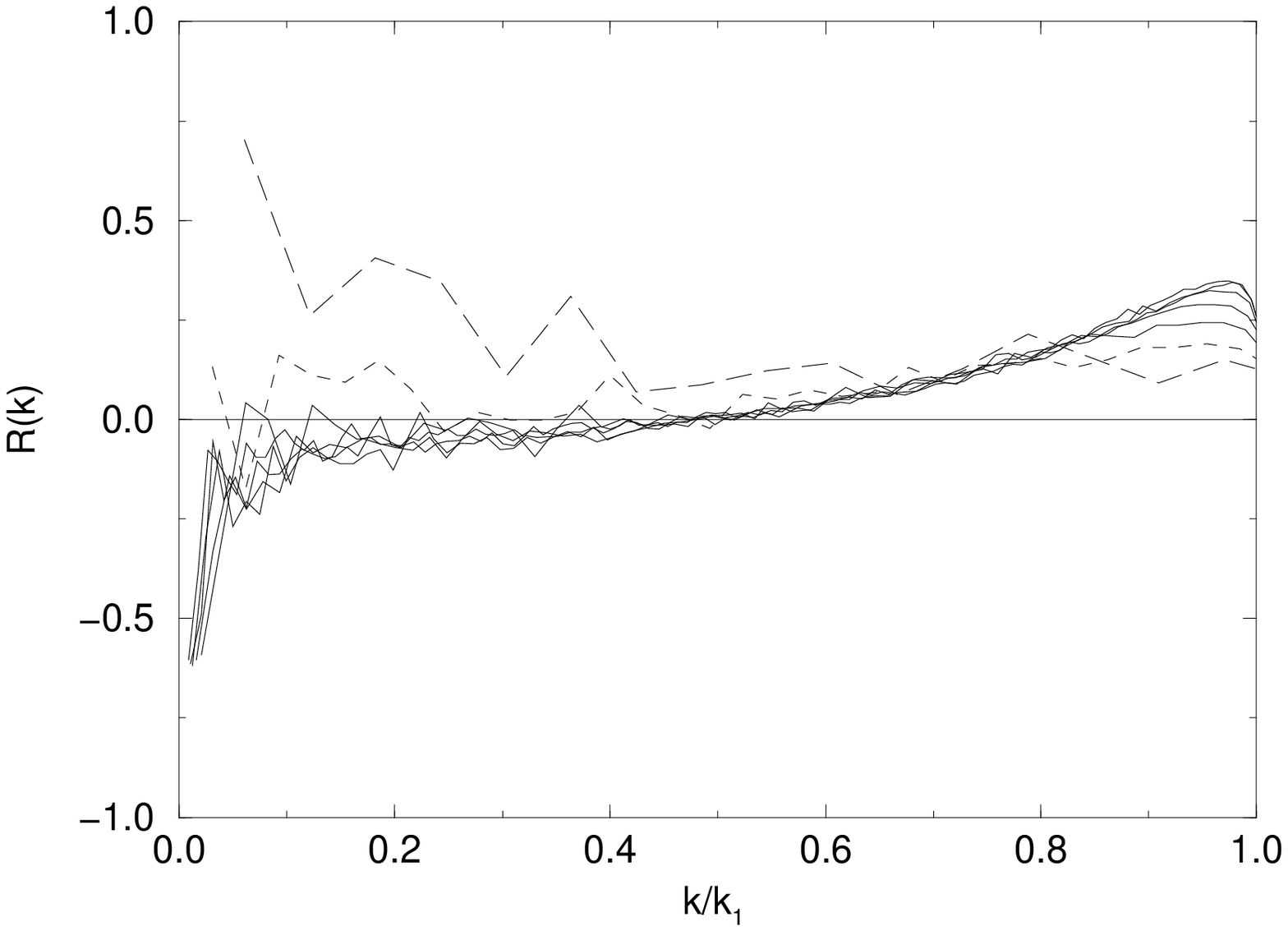,width=8truecm}}
\caption[]{\small\sf
         {\bf Phase-correlation.}\\
         $R(\{\psi^{(--)}_<\}_\theta,\{\tilde{\psi}_<^{(--)}\}_\theta;k)$
         for cutoff wavenumbers
         $k_1=16.5$ (\thinlongdashed), $32.5$ (\thindashed),
         $48.5, 64.5, 80.5, 96.5, 112.5$ (\thinline) with
	 $k_0=128$.
        }
\label{evcorr_arg fig}
\end{figure}

\begin{figure}
\centerline{\psfig{figure=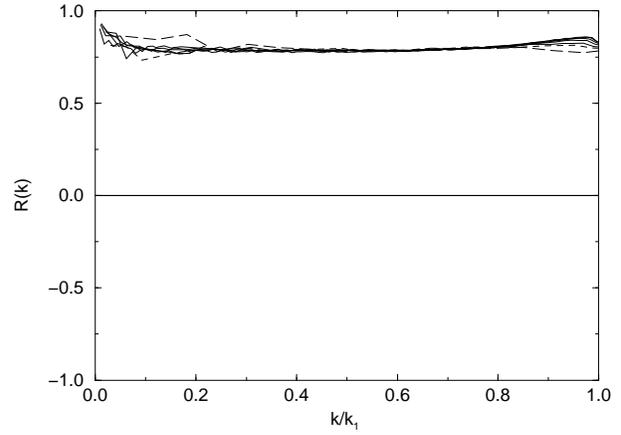,width=8truecm}}
\caption[]{\small\sf
         {\bf Amplitude-correlation.}\\
         $R(\{\psi^{(--)}_<\}_r,\{\tilde{\psi}_<^{(--)}\}_r;)$
         for cutoff wavenumbers
         $k_1=16.5$ (\thinlongdashed), $32.5$ (\thindashed),
         $48.5, 64.5, 80.5, 96.5, 112.5$ (\thinline) with
	 $k_0=128$.
        }
\label{evcorr_mag fig}
\end{figure}

In Figure \ref{evcorr_arg fig} we present the correlation
obtained from phase information only and see that the general
picture is the same as when we considered the whole fields.
In Figure \ref{evcorr_mag fig}, however, where we consider
only amplitude information, we see that there is uniformly
good correlation at a level of around $0.75$.

This means that the poor correlation between the exact
subgrid terms and an eddy-viscosity model is due almost
entirely to mismatching phases.  However, by choosing
$\delta\nu (k)$ suitably, it is possible to provide a good
match as far as amplitude is concerned.

\section{CONCLUSION}

We have seen that, for the range of cutoff wavenumbers considered
here, there is some degree of universal behaviour when we consider
the partitions of the nonlinear term.  We see
that for $k<k_1$, the terms involving coupling between low-wavenumber
modes dominate, and this is a useful property when considering the
potential for large eddy simulations.
We have also seen that while it is not possible to represent
subgrid terms exactly using an eddy-viscosity model, it is
possible to model the {\sl amplitude} of these
terms.  This will lead to reasonable results when one
considers, for example, the flow of energy but it is
clear that on a more fundamental level the velocity field
will become corrupted due to phase-errors.

\vspace{0.5truecm}
\noindent{\sl Acknowledgements:}
The results presented in this paper have been
obtained from simulations
performed on the computers of the Edinburgh
Parallel Computing Centre.  The research of
A. Young is supported by the
Engineering and Physical Sciences Research Council.

\section{REFERENCES}

Kraichnan, R.H., ``Eddy viscosity in two and three dimensions,''
{\sl J. Atmos. Sci.}, Vol. 33, pp. 1521--1536.

Lesieur, M. and Rogallo, R., 1989,
``Large-eddy simulation of passive scalar diffusion in isotropic
turbulence,''
{\sl Phys. Fluids A}, Vol. 1, No. 4, pp. 718--722.

McComb, W.D., 1995, ``Theory of turbulence'',
{\sl Rep. Prog. Phys.}, Vol. 58, pp. 1117-1206.

McComb, W. D., Roberts, W. and Watt, A. G., 1992, ``Conditional-averaging
procedure for problems with mode-mode coupling'', {\sl Phys. Rev. A},
Vol. 45, pp. 3507--3515.

McComb, W. D. and Watt, A.G., 1992, ``Two-field theory of
incompressible-fluid turbulence'', {\sl Phys. Rev. A}, Vol. 46,
pp. 4797--4812.

McComb, W.D., Yang, T.-J., Young, A.J. and
Machiels, L., 1997, ``Investigation of renormalization
group methods for the numerical simulation of isotropic
turbulence,'' {\sl Proc. 11th Symposium on Turbulent
Shear Flows, Grenoble}, pp. 4-23--4-27.

Machiels, L., 1997, ``Predictability of small-scale motion in
isotropic fluid turbulence,'' {\sl Phys. Rev. Lett.}, Vol. 79,
No. 18, pp. 3411--3414.

Orszag, S., 1969, ``Numerical Methods for the Simulation of Turbulence,''
{\sl Phys. Fluids (suppl. 2)}, Vol. 12, pp. 250--257.

Orszag, S., 1971, ``Numerical Simulation of Incompressible Flows Within
Simple Boundaries.  I. Galerkin (Spectral) Representations,''
{\sl Stud. Appl. Maths.}, Vol. 50, No. 4, pp. 293--327.

Rogallo, R.S., 1981, ``Numerical Experiments in Homogeneous Turbulence,''
NASA TM-81315.

\end{document}